\documentclass[11pt,a4paper]{article}
\usepackage{amsmath,amssymb,amsfonts,a4wide,graphicx,bm,times}
\usepackage[colorlinks=true,linkcolor=black, citecolor=black,
urlcolor=black]{hyperref} \input{epsf}
\numberwithin{equation}{section}
\makeatletter \let\old@startsection=\@startsection
\renewcommand{\@startsection}[6]
{\old@startsection{#1}{#2}{#3}{#4}{#5}{#6\mathversion{bold}}}
\makeatother
\def\hb{\hbar }
\usepackage[utf8]{inputenc}
\def\gg{{ G}}
\def\Tr{{\rm tr\, }}
\def\defeq{\stackrel{\text{def}}=}
\newcommand\re[1]{({\ref{#1}})}
\def\be{\begin{eqnarray}}
    \def\ee{\end{eqnarray}}
    \def\no{\nonumber}
    \def\la{\label}
    \def\l {\l}
\def\({\left(} \def\){\right)} 
\def\<{\left\langle\,} 
\def\>{\, \right\rangle} 
\def\[{\left[}
 \def\]{\right]} 
 \def\cl{ {\rm cl}}
    \def\hf{ {\textstyle{1\over 2}} }

\def\d{\delta} 

  \def\p{\partial}
  \def\a{\alpha}
 \def\b{\beta}

  \def\t{\tau}

\def\CA{{\cal A}}\def\CB{{\cal B}} 
  \def\CF{{\cal F}}
 
 \def\reg{{\rm reg}}
 \def\O{\Omega} 

\def\CN{{\cal N}}
\def\CZ{{\cal Z}}

\def\l{z}
\def\CA{{\cal A}}

\def\p{\partial}

\def\IZ{ {\mathbb Z}}
\def\IC{ {\mathbb C}}

\def\o{ \omega }

  \def\CE{{\cal E}}

\def\w{ w }
\def\nb{ n_{_B}}
\def\qu{ \text{qu}} \def\gauss{ \text{gauss}}

\def\tw {\text{tw}}

\def\netwo{ { \text{ odd} } }
 \def\br{\begin{remark}\rm\small}
\def\er{\end{remark}}
\newtheorem{remark}{Remark}[section]

\newcommand{\virg}{{\qquad , \qquad}}

\newcommand{\Res}{\mathop{\,\rm Res\,}}
\newcommand{\beq}{\begin{equation}}
\newcommand{\eeq}{\end{equation}}
\newcommand{\bea}{\begin{eqnarray}}
\newcommand{\eea}{\end{eqnarray}}

\def\bd{\begin{definition}}
\def\ed{\end{definition}}
\newtheorem{definition}{Definition}[section]

\begin{document}

\thispagestyle{empty}

\begin{flushright}
IPhT-t10/077\\
CERN-PH-TH-2010-128\\
\end{flushright}

\vspace{1cm}
\setcounter{footnote}{0}

\begin{center}

 {\Large\bf CFT and topological recursion}

\vspace{20mm} 

Ivan Kostov$^{\star}$ \footnote{\it Associate member of the Institute
for Nuclear Research and Nuclear Energy, Bulgarian Academy of
Sciences, 72 Tsarigradsko Chauss\'ee, 1784 Sofia, Bulgaria} and
Nicolas Orantin$^{\dag}$ \\[7mm] \vskip2mm {\it $^{\star}$ Institut de
Physique Th\'eorique, CNRS-URA 2306 \\
	     C.E.A.-Saclay, \\
	     F-91191 Gif-sur-Yvette, France \\[5mm] }

{\it $^{\dag}$
Theory division, CERN\\
CH-1211 Geneva 23, Switzerland.  }
 \end{center}

\vskip18mm

 \noindent{ We study the quasiclassical expansion associated with a
 complex curve.  In a more specific context this is the $1/N$
 expansion in $U(N)$-invariant matrix integrals.  We compare two
 approaches, the CFT approach and the topological recursion, and show
 their equivalence.  The CFT approach reformulates the problem in
 terms of a conformal field theory on a Riemann surface, while the
 topological recursion is based on a recurrence equation for the
 observables representing symplectic invariants on the complex curve.
 The two approaches lead to two different graph expansions, one of
 which can be obtained as a partial resummation of the other.  }

\newpage

\setcounter{page}{1}
  
\section{Introduction}
\label{sec:Introduction}

The $1/N$, or topological,  expansion in $U(N)$ invariant
matrix models is being considered with
renewed interest in the last years because of its various applications
in topological string theories on special classes of Calabi-Yau
geometries or $\CN=2$ superconformal gauge theories 
\cite{Dijkgraaf:2002fc,Bouchard:2007ys,Dijkgraaf:2009aa}.
The solution in the limit $N\to\infty$ is generically 
described by a { complex curve}.  This is so also in the case when the
symmetry group is a direct product of several unitary groups, as the
ADE matrix chains \cite{Kostov:1989eg, Kostov:1995xw}.  Since the
complex curve determines the classical spectral density of the matrix
variables, it is also called {\it spectral curve}.
In all solved examples, the $1/N$ expansion can be formulated entirely
in terms of the complex curve.  The fact that the spectral curve
determines not only the classical limit $N\to\infty$, but also the
complete $1/N$ expansion, is highly non-trivial and deserves to be
better understood.
   
Two approaches to the $1/N$ expansion formulated in terms of the
spectral curve were developed in the past ten years.  The first
approach, which will be refered to as CFT approach, is based on the
{\it conformal invariance}, which is believed to be present in all
$U(N)$ invariant matrix systems.  The conformal symmetry is obvious
for the class of matrix models which, after diagonalization, reduce to
Coulomb gases \cite{Marshakov:1991gc,Kharchev:1992iv, Kostov:1999xi}.
A prescription to evaluate the quasiclassical expansion in such matrix
models using the toolbox of CFT was outlined by one of the authors in
\cite{Kostov:1999xi, MMCFT-LesHouches}.  The basic idea is that the
conformal invariance is sufficient to construct the collective field
to all orders in the quasiclassical expansion.  The spectral curve
appears as the classical value of a bosonic field, which lives on a
Riemann surface associated with this spectral curve.  The gaussian
approximation gives the two leading terms of the $1/N$ expansion,
while the higher terms are obtained by inserting special local
operators at the {\it branch points} of the Riemann
surface.\footnote{The field-theoretical description of the branch
points was pioneered by Al.  Zamolodchikov
\cite{Zamolodchikov:1987ae}.} 
The proposal of
\cite{Kostov:1999xi, MMCFT-LesHouches} is a field-theoretical
formulation of the `method of moments' for performing the higher genus
calculations, developped in \cite{Ambjorn:1992gw}, and generalizes the
CFT description of the non-critical string theories
\cite{Dijkgraaf:1990rs, fukuma1991continuum, verlinde1991solution}.
The case of a hyperelliptic Riemann surfaces was considered in details
in \cite{Kostov:2009aa}.

Another approach to the $1/N$ expansion, known as {\it topological
recursion}, was developed more recently by B. Eynard and collaborators
\cite{Eynard:2004mh, Eynard:2005aa, Chekhov:2005rr, Chekhov:2006vd,
Chekhov:2006rq, Eynard-formal06, Eynard:2007kz, Eynard:2009zz,
borot-2009, Eynard:2008he, 1751-8121-42-29-293001}.  Whilst the CFT
approach is related to field theory, the topological recursion, which
we denote shortly by TR, is related to algebraic geometry.  The
topological recursion gives a very efficient algorithm for calculating
the $1/N$ expansion of the observables representing symplectic
invariants on the complex curve.  The recursion procedure overcomes
some of the technical difficulties of the method of moments
\cite{Ambjorn:1992gw}.  The basic observation here is that recurrence
equation is most simply formulated in terms of residues at the branch
points of the Riemann surface.  The outcome of the topological
recursion is an elegant graphical scheme, containing only trivalent
vertices.

Let us emphasize  that the CFT and TR approaches  can 
be  formulated only in 
terms of the spectral curve, without any reference to  the underlying 
matrix model. Therefore, instead of speaking of  the $1/N$ expansion, 
we will speak of the quasiclassical expansion  
 (in $\hbar = 1/N$) associated with the spectral curve.
Since any spectral curve can be considered as a
solution of the universal Whitham hierarchy \cite{krichever1992tau},
 CFT and TR  give two methods to evaluate the quasiclassical expansion
of an integrable hierarchy, given its dispersionless  limit.\footnote{
Reconstruction of the quasiclassical expansion for the KP hierarchy,
named $\hbar$-dependent KP hierarchy, was considered in
\cite{Takasaki:1995aa, Takasaki:2009aa}.  A serious problem in these
works, which are based on the Lax formalism, is that the recursion
relations are extremely complicated.  The advantage of the CFT/TR
method, based on the Virasoro symmetry, is that the solution is
obtained in a simple and easy to manipulate form. We are not
going to discuss here the uniqueness of the quasiclassical expansion.
CFT/TR is believed to give the same quasiclassical expansion as
the Lax formalism, but as far as we know there is no general proof of
that.}  
  %

The aim of this paper is to compare the two methods and the diagram
techniques they lead to.  Since the CFT method and the topological
recursion represent two different techniques to resolve the Virasoro
constraints, they are expected to lead to the same solution, although 
this is far from obvious for  general spectral curve.
On the other hand, the two diagram techniques are seemingly very
different and have different geometrical interpretation in terms of 
world sheets.  Therefore we think that it is important and
instructive to find out the exact correspondence between the two forms
of the quasiclassical expansion associated with a spectral curve.  
In this paper we show that the two diagram
techniques lead to the same result and that moreover one of them is
obtained by a partial resummation of the other.  We conclude that the
CFT method and the topological recursion are two different
realizations of the same procedure, which we will refer to as CFT/TR.
We underline that we are considering the case of the most general, not
necessarily hyperelliptic, compact complex curve\footnote{From this perspective, the result of this 
paper goes beyond the unitary one matrix model. It applies for example ti the unitary two matrix 
models where the spectral curve is not hyperelliptic in general.}.


 The paper is organized as follows. In Sect. 2 we collect  some  definitions
 and notations related to the  spectral curve.  In Sect. 3 we describe the 
 CFT approach. We first remind the operator formalism for a
 gaussian field on a Riemann surface. Then we express the 
 Virasoro constraints  the operator formalism for a gaussian 
 field on the spectral curve.  We  write  the Virasoro constraints 
 in terms of the mode expansion of the gaussian field near the 
 branch points and construct the dressing operators to be inserted at 
 the branch points in terms of these modes. Finally we derive the 
 diagram technique for computing the higher genus free 
   energies and correlation functions.   In Sect. 4 we first give  a brief 
   summary of the topological recursion.  Then we rewrite the equations
   of TR in terms of the  series expansions of the symplectic invariants 
   near the branch points.  We start with the simplest case of a spectral curve 
   with only one,  simple,  branch point.  In this case the symplectic invariants
    are given by the observables of the Kontsevich model.  We
    formulate the equations of  TR  in terms of  a trivalent 
    graph expansion. After that  we consider the case of an arbitrary 
    spectral curve.   Sect. 5 we give the exact correspondence between the
     diagram techniques in the CFT and the TR approaches.

 \section{Spectral curve and classical free energy}

  By spectral curve we understand the triple
\be\la{defSC} {\cal E}\defeq (\Sigma ,x,y) , \ee
where $\Sigma $ is a Riemann surface of genus $ \gg $ and the
functions $x$ and $y$ are analytic in some open domain of $\Sigma $.
The Riemann surface is assumed to have $l$  punctures ($l\ge 1$), 
where the differential
$ydx$ becomes singular.  
For later convenience we introduce a global parametrization $z$ of the
Riemann surface  $\Sigma $.  The complex variable $z$ belongs either to
the Riemann sphere (for genus 0) or to a quotient of the unit disk
(for genus 1 or larger).  In terms of the global parameter $z$ we have
\beq x=x(z), \ \ y = y(z)\, , \eeq
where $y$ and $x$ are analytic functions of $z$.  
In the  neighborhood of the punctures $z=\hat a_\a $
  we chose local coordinates $\zeta_{(z,\alpha)} $ such that  
  \be
  \zeta^{-1}_{(\hat a_\a,\alpha)} =0 \quad (a=1,\dots, l). 
  \ee
  We can turn the spectral 
  curve into a set of algebro-geometric data for Krichever's universal Whitham
hierarchy \cite{krichever1992tau}.  
We introduce the classical collective field as the one-form
\be \la{defPhicl}
d S(z) =  y(z) d x(z), \ee
\def\nfty{ n_{\infty}}
which is by definition holomorphic on the Riemann surface except at
the punctures $z=\hat a_\a$.   
 The function $S(x)$ is nothing but the potential for
the universal Whitham hierarchy.  
The classical free energy associated with the spectral curve is
defined as \cite{krichever1992tau}
\be
\CF^{(0)}[S]
= \int _{\Sigma} \bar d S\wedge dS.
\la{deffren}
\ee

The deformations of the spectral
curve can be represented as commuting flows in the phase space
corresponding to the ``times'' associated with the punctures.  The Poisson
brackets defined in the space  of functions of the two variables $x$ and $y$ are defined as
\be
\{ f, g\} \defeq \p_x f \, \p_y g - \p_x g \, \p_y f
\la{poissbk}
\ee
and correspond to the symplectic form $\o = dx \wedge dy$.
 The classical field $S$ can be thought of as the action of a Hamiltonian
system depending on the coordinate $x$ and the ``times" associated
with the expansion at the punctures \cite{krichever1992tau}.

The classical problem in the universal Whitham hierarchy is to
determine the one-form \re{defPhicl} by its asymptotics at the
punctures and the moduli associated with the non-contractible cycles.
The moduli of the spectral curve
at the punctures are given by the principal parts
\be t_{\alpha} \equiv
\Res_{z \to \hat a_\alpha} y(z) dx(z) \ee
 and the potentials 
 \be
V_{\alpha}(z) \equiv \Res_{z' \to \hat a_\alpha} y(z') dx(z') \ln \left( 1
- {\zeta_{(z,\alpha)} \over \zeta_{(z',\alpha)}}\right) .
\ee
%
%
Let $\{ \CA_i, \CB_i\}_{i=1}^G$ be a canonical basis of cycles on
$\Sigma$.
The classical solution is completely determined by its asymptotics
at the punctures and the   `filling fractions'
 \be\la{normMC} \nu_k = {1\over 2\pi i} \oint\limits _{\CA_k} y(z)
 dx(z) \qquad (k=1,\dots, G) \ee
 associated with the A-cycles.  A basis of holomorphic one-forms on
 the Riemann surface associated with the cycles $\CA_k$ is given by
 the derivatives
  \be\la{abdIf} \o_k = { \p \over\p {\nu _k} } y dx .  \ee
Their integrals along the A- and B-cycles are given by
\be\la{eeqa} \frac{1}{2\pi i}\oint_{\CA_k} \o_l = \delta_{kl} , \qquad
{1\over 2\pi i} \oint_{\CB_k} \o_l=\tau_{kl} , \ee
where $\t=\{\t_{ik }\}$ is the period matrix of the Riemann surface.
We will consider the filling fractions as a set of external
parameters, which are part of the moduli of the spectral curve.

 An important role in the construction of the quasiclassical expansion 
 is played by the {\it  Bergman kernel} for the spectral curve,
 $B(z,z')$.  The Bergman kernel  is defined uniquely
by the following three conditions:

-- it is globally defined on the Riemann surface, including the
punctures;

-- it behaves at small distances as
\be\la{twoptf} B(z,z') = {dz \, d z'\over (z-z')^2} + \ \text{regular
function}; \ee

  -- it has has vanishing integrals around the $\CA$-cycles: \be
  \oint_{{\cal A}_i } B(z,z')= 0, \qquad (j=1, \dots, G).  \ee
The Bergman kernel \re{twoptf} depends on the spectral curve only
through the $3G-3+l$ (for $G\ge 2$) complex moduli of the punctured
Riemann surface $\Sigma$.  It is important noticing that it does not
depend on the moduli associated with the singular behavior of $ydx$
near the punctures.

Below  we will  assume that the classical problem is solved and the one-form
\re{defPhicl} is already known.  The free energy \re{deffren}, or the 
classical action for the collective field, can be written as a sum of residues
associated with the punctures and contour integrals of the one-form $ydx$
\cite{krichever1992tau, Bertola:2003rp,
{Chekhov:2006vd}}. For the sake of completeness  we remind the 
general expression for $ \CF^{(0)}[S] $, although we will not  use it:
   %
  \be
   \CF^{(0)}[S] =
    - {1\over 2  }
  \[
   \sum_{\alpha} \Res_{z \to \hat a_\alpha} V_{\alpha}(z) \, y(z) dx(z)
 +  \sum_\alpha t_{\alpha} \mu_{\alpha}
 +  \sum_{i =1}^G 
   \nu_i    \oint \limits_{\CB_i} y(z) d x(z) \!  \], 
  \ee
  where 
\be \mu_\alpha \equiv \int_{(z =\hat a_\alpha)}^o\left( y(x)dx(x) -
dV_\alpha(z) + t_\alpha {d\zeta_{(z,\alpha)} \over
\zeta_{(z,\alpha)}}\right) + V_\alpha(o) - t_\alpha
\ln(\zeta_{(o,\alpha)}) \ee
for an arbitrary base point $o$.  

 \section{The CFT approach}
\label{sec:CFT}

 The CFT approach is based on the assumption
that the quantum collective field, which we denote by  $\Phi$,
 has the same analytical properties as 
the classical solution $S(x)$. In other words,  the field $\Phi$ 
 is invariant with respect to the
conformal transformations of the spectral parameter $x$ that preserve
the $l$ punctures $\hat x_\a =x(\hat a_\a)$ of the Riemann 
surface.\footnote{The invariance with
respect to conformal transformations  which do not preserve the punctures 
determines the reaction of the system to a change of the external
parameters.  This symmetry is not relevant for the quasiclassical expansion.  }
The outcome of the CFT approach is an universal formula for the 
quasiclassical expansion of the free energy 
  \be \CF   [S, \hb ] = \sum_{g\ge 0} \hb^{2g-2}\
  \CF^{(g)}[S] + \text{non-perturbative terms}. \ee
 in terms of the  classical solution $S(x)$.

\subsection{Gaussian field on the spectral curve}

In the gaussian approximation, the classical field $\Phi_\cl$ is the
 expectation value of a gaussian field $\Phi$.  We thus split
the collective field into classical and quantum parts:
\be \Phi = \Phi_\cl +\Phi_\qu, 
\qquad   \Phi_\cl  = {1\over \hbar} S. 
\la{phicphiq}
 \ee
The subleading term in the quasiclassical  expansion for the free energy,
$\CF_\gauss= \CF^{(1)}$, comes from the gaussian fluctuations around
the classical solution.  The corresponding factor in the partition
function is given by the inverse power of the holomorphic piece of the
determinant of the Laplace operator on the Riemann surface:
 \be \CZ_\gauss = {1\over \det \bar \p_0} .  \ee
The expression of the chiral determinant for a hyperelliptic curve has
been given in \cite{Zamolodchikov:1987ae} and for a general surface in
\cite{Verlinde1987357, Knizhnik:1989ak}.

All correlation functions in the gaussian approximation can be
expressed through the two-point function of a gaussian field on the
Riemann surface,  which is given by the Bergmann kernel,
\be\la{defBb} B(z,z') = \< d\Phi_\qu(z) d\Phi_\qu(z') \>_{_\Sigma}. \ee
%
%
%
%
%

  \subsection{Operator formalism for the gaussian field on a Riemann
  surface}

Let $z_0$ be a base point on the Riemann surface $\Sigma$ and
$\xi_{(z)}$ be a  local coordinate  in the  neighborhood  of
$z_0$ such that $\xi_{(z_0)}=0$.  We can associate with this variable
a Hilbert space by expanding the gaussian field in Laurent series.  We
split the quantum field into a singular and a regular parts
   \def\sing{ { ^{\text{sing}}}} \def\reg{ {^{\text{reg}}}}
\be d\Phi _\qu (z) = d\Phi _\qu (z)^\sing + d\Phi_\qu (z) ^\reg, \ee
having mode expansions
\be\la{modexpPhi} d\Phi_\qu (z) ^\sing = \sum_{ n\ge 0 } J_n \,
\xi_{(z)} ^{-n-1} d \xi_{(z)} , \quad d\Phi(z) _\qu^\reg= \sum_{ n\ge
1 } J_n \, \xi_{(z)} ^{n-1} d \xi_{(z)} .  \ee The amplitudes $J_n$
are assumed to satisfy canonical commutation relations \be [J_m , J_n
]= m\, \d_{m+n, 0}, \qquad m,n\in\IZ. \ee
This mode expansion defines a Hilbert space with left and right Fock
vacua satisfying \footnote{The amplitude $J_0$ commutes with the rest
and its action on the vacuum states can be specified separately.  We
can define a vacuum state with charge $N$ by requiring that
$J_0|0\rangle = N|0\rangle$.  Alternatively one can introduce the zero
mode $\Phi_0$, which is canonically conjugated to $J_0$, so that $ [J
_0, \Phi_0]= 1$.  Then the vacuum state with charge $N$ is $ e^{
N\Phi_0} |0\rangle$.  }
\be \langle 0 | J _{-n} = 0, \quad J _n |0 \rangle = 0 \qquad (n\ge
1).  \ee
From the commutation relation
 \be [ d\Phi _\qu(z) ^\sing, d\Phi_\qu ^\reg (z') ] = {d\xi_{(z)}
 d\xi_{(z')}\over [\xi_{(z)} - \xi_{(z')}]^2}, \ee
 one finds for the correlation function
 \be\la{twobare} \langle 0 |d \Phi_\qu (z) d\Phi_\qu (z')|0\rangle=
 {d\xi_{(z)} d\xi_{(z')}\over [\xi_{(z)} - \xi_{(z')}]^2} .  \ee

The left and right Fock vacua defined above are such that the gaussian
field $\Phi_\qu$  has vanishing expectation value and its two-point function
\re{twobare} gives the singular part of the Bergman kernel.  We would
like to deform the left and right vacuum states in such a way that the
 two-point function of $\Phi_\qu$ reproduces also the regular part of the Bergman kernel \re{twoptf}.  In order to obtain the operator representation 
 of the  collective field $\Phi$,  eqn. \re{phicphiq}, we should also introduce the 
 expectation  expectation value $\Phi_\cl $. 
  For that we first expand the classical field and the Bergman kernel in the
local variable $\xi_{(z)} $,  assuming  that  $\Phi_\cl^\sing (z_0)=0$,
 i.e.  $z_0$ is a regular point on the Riemann surface:
\be\la{expPhic} d\Phi_\cl(z) &=& {1\over \hbar}
 \sum_{ n\ge 1} \mu _n \ \xi_{(z)}
^{n-1} d \xi_{(z)} , \\
B(z,z') &=& {d\xi_{(z)} d\xi_{(z')}\over [\xi_{(z)} - \xi_{(z')}]^2} +
\sum_{ n , m\ge 1} B_{m, n} \ \xi_{(z)} ^{n-1}\ \xi_{(z')} ^{m-1} \ d
\xi_{(z)} \, d \xi_{(z')} .  \ee
Then, as suggested in \cite{Vafa:1987es, Matsuo:1986ph,
Dijkgraaf:1987vp}, we Bogolyubov transform   the left Fock vacuum,
\be\la{defVac} \langle 0 | &\to& \langle \Sigma |= \langle 0 | \exp\(
\hf \sum _{m, n\ge 1} {B_{m, n} \over m\, n}J_m J_n+ {1\over \hbar}
\sum_{n\ge 1} {\mu_ {n}\over n} J_{n}\).  \ee
In this way the Riemann surface $\Sigma$ is represented by 
a state  $\langle \Sigma|$  belonging to the Fock space associated with the base point
$z_0$. 
For any correlation function of $d\Phi_\qu(z_i)$, with $z_i$ belonging
to a coordinate patch which contains the point $z_0$, the unnormalized
expectation value on the surface $\Sigma$ is given by
 \be\la{defSigma} \langle d\Phi_\qu(z_1) \dots d\Phi_\qu(z_n)
 \rangle_{_\Sigma} \ =\ \langle \Sigma | d\Phi_\qu(z_1) \dots
 d\Phi_\qu(z_n) | 0 \rangle .  \ee

If we need to evaluate the correlation functions of operators that
belong to different coordinate patches of $\Sigma$, we should
generalize the operator representation for a Riemann surface with
several extra punctures $z_1, \dots, z_r$.  To each puncture we
associate a pair of vacuum states as above and define the right and
the left Fock vacua as the direct products of the Fock vacua
associated with each puncture.  Let $\xi_{(z,s)} $ be the local
coordinate in the $s$-th coordinate patch.  Then \re{defVac}  
generalizes to
\be\la{defVacm} \langle 0 | =\bigotimes_{s=1}^r \langle 0 ^{(s)} |
&\to& \langle \Sigma |= \langle 0 |\exp\(\hf \sum_{m,n\ge 1} {1\over
mn}B_{m, n}^{(s,s')} J_m^{(s)} J_n^{(s')} +{1\over \hbar}
 \sum_{n\ge 1} {1\over n}\mu^{(s)} _ {n}   J^{(s)} _{n}\), 
  \ee
 where
 \be 
 \no 
 \mu_n ^{(s)} &=& \Res_{z\to z_s}  y(z)d x(z) \ \xi_{(z,s)} ^{-n},
 \la{defmus} \\ \no B_{m,n}^{(s,s')}&=& \Res_{z\to z_s} \Res_{z\to
 z_{s'}} B(z,z') \ \xi_{(z,s)} ^{-m} \xi_{(z', s')} ^{-n} .
 \la{defBmn}\ee

\subsection{Conformal invariance at  the branch points}

The stress-energy tensor for the conformal transformations of the
global parameter $z$ is
   \be\la{defTz} T (z) dz^2={1\over 2} \lim\limits_{z'\to z} \[
   d\Phi(z) d\Phi(z') - B(z,z')\] .  \ee
The gaussian field is by definition   invariant under such transformations
 in the sense that the
expectation value $\< T(z)\> _{_\Sigma}$ is analytic everywhere except
at the punctures $z=\hat a_\a$ where the classical solution has poles.

 The basic assumption of the CFT approach is that the theory is
 invariant with respect to
 conformal transformations of the spectral variable $x$. 
 The
 conformal invariance in the $x$-plane leads to a stronger condition,
 \be\la{confWI} \({dz\over dx}\)^2\< T (z) \> _{ _\Sigma}=
 \text{analytic function of  } x  
 \ee
with possible singularities at the punctures $\hat x_\a= x(\hat a_\a)$.  The
condition \re{confWI} is satisfied everywhere on the Riemann surface
except at the branch points\footnote{Strictly speaking, we should use
the term ramification points instead of branch points.  According to
the standard terminology, the points $a_s\in \Sigma$ where $dx(z)=0$
are ramification points while the points $x(a_s)\in\IC $ are branch
points.  Each branch point is the image of one or more ramification
points in the $x$-plane.} where $\p_z x=0$.  This means that the
gaussian field is not a good approximation near the branch points.

 If the vicinity of the branch point $a_s$ one can define an involution
$z\leftrightarrow \tilde z $, such that $\tilde z \ne z$ and $x(\tilde
z )=x(z)$.  Up to quadratic terms in $z-a_s$, $\tilde z-a_s= a_s-z$.
In general, the involution $z\leftrightarrow \tilde z $ depends on the
branch point; it is globally defined only if the spectral curve is
hyperelliptic.  The Hilbert space associated with base point $z=a_s$
splits into odd and even sectors with respect to the involution
$z\leftrightarrow \tilde z$. 
If the Riemann surface $\Sigma$ is viewed as a branched cover of the
spectral plane $x$, the involution $z\leftrightarrow \tilde z$ is the
monodromy around the branch point $x_s = x(a_s)$.  This is why the odd
sector will be called {\it twisted sector}.\footnote{ The richer
structure of the Hilbert space at the branch points is related to the
extended symmetry.  For a branch point this is the $\hat u(2) = \hat
su(2)\oplus \hat u(1) $ current algebra.  In the case of a branch
point of order $m$ the symmetry is $\hat u(m)$.  The Cartan subalgebra
of $\hat u(m)$ is spanned by the $m$ currents associated with the
different sheets of the Riemann surface near the branch point.}

 Each sector is characterized by its
stress-energy tensor. The stress-energy tensor for the even sector 
automatically  sarisfies \re{confWI}  near the branch point, while the  
stress-energy tensor for the odd sector does not.
We
define the twisted component of the collective field as the projection
    \be\la{defPhis}
     \Phi ^{[s]}(z) \ \defeq\ {\Phi ( z) - \Phi (\tilde z)\over 2}
    \qquad \text{for} \ z \ \text{in the vicinity of}\ a_s.  \ee
The two-point function and the stress-energy tensor of the twisted
field are\footnote{Since the function is symmetric with respect to exchanging 
  the two arguments, one can antisymmetrize only with respect of the second argument.}
\be\la{deftwcor} \< d\Phi^{[s]}_\qu(z)\
d\Phi^{[s]}_\qu(z')\>_{_\Sigma} =
B^{[s]}(z,z')\equiv \hf B(z,z') - \hf B(z, \tilde z'),
\ee
\be\la{defTztw} T ^{[s]}(z) dz^2&=& {1\over 2} \lim\limits_{z'\to z}
\[ d\Phi^{[s]}(z) d\Phi^{[s]}(z') -  B^{[s]} (z,z')  \] .  \ee
   %
The condition \re{confWI} represents a non-trivial constraint for the twisted 
stress-energy tensor  $T ^{[s]}(z) $.

We assume that all branch points are simple.  Moreover, we assume that
these points are neither poles of $y$ neither zeroes of $dy$.  If
$a_s$ is one of the branch points, i.e. $dx/dz \sim (z-a_s)$,
\re{confWI} implies that the Laurent expansion of the stress-energy
tensor contains only even powers of $z-a_s$,
\be \la{Tzeven}
T^{[s]} (z)= \sum _{n\in\IZ} L^{[s]}_{2n}\, (z- a_s)^{-2n-2}, \ee
and the non-vanishing  Virasoro operators $L_{2n}$ satisfy
\be
\<L^{[s]}_{2n} \>_{_\Sigma} =0, \qquad n\ge - 1.
\ee

 \subsection{Dressing operators}
 
Our aim is to construct dressing operators located at the branch
points, which restore the conformal invariance.  For that we will apply
the operator formulation \re{defVacm}--\re{defmus} to the special case
when the extra punctures are placed at the branch points $z=a_s \
(s=1, \dots , \nb)$.
    
Given a local coordinate $\xi _{(z, s)}$, such that $\xi _{(a_s,
s)}=0$, the mode expansion for the twisted field \re{defPhis} is given
by the odd part of the mode expansion \re{modexpPhi},
\be
d\Phi^{[s]}_\qu (z) &=&   \sum_{p\netwo } J_p^{[s]} \ \xi _{(z, s)} 
^{- p -1} d\xi _{(z, s)}  ,
 \no \\
\la{modexJ} d\Phi_\cl^{[s]}(z) &=& {1\over \hbar} \sum_{ p\ge 3,
\netwo } \mu _p^{[s]} \ \xi_{(z,s)} ^{p-1} d \xi_{(z,s)} .\ee
We denote by $|0^{[s]}_\tw \rangle$ the twisted Fock vacuum associated
with the branch point $a_s$.  From the point of view of a CFT on the
spectral plane, this state can be considered as the result of the
insertion at the point $x_s$ of a twist operator of conformal weight
${1\over 16}$.  The Hilbert space space associated with the ensemble
of the branch points of the Riemann surface is spanned by the states
of the form
    \be \langle 0_\tw | \prod_{i} J_{p_i}^{[s_i]} , \qquad \prod_{i}
    J_{-p_i}^{[s_i]} |0_\tw \rangle \qquad (p_i \ge 1, \netwo), \ee
where the left and right Fock vacua are defined as
      \be \langle 0 _\tw | = \bigotimes_{s=1}^{\nb} \langle0 _\tw
      ^{[s]}|, \qquad |0 _\tw \rangle = \bigotimes_{s=1}^{\nb} |0 _\tw
      ^{[s]}\rangle .  \ee

Once we have an operator representation of the expectation value in
\re{confWI}, the construction of the dressing operators  becomes a
purely algebraic problem.  The result does not depend on the choice of
 parametrization, but the mode expansion \re{modexJ} does.  Let us
pick a canonical parametrization in the vicinity of each branch point.
We would like to choose the local parametrization variable $ \xi_{(z,
s)}$ so that the moments $\mu_p^{[s]}$ are simply expressed in terms
of the moduli $ M^{[s]}_p$ of the spectral curve, defined
as\footnote{These moduli are linear combinations of the ACKM moments
\cite{Ambjorn:1992gw}.}
 \be M^{[s]}_p= \oint\limits _{a_s} {y (z) dx(z)\over
 [x(z)-x(a_s)]^{p/2}}, \qquad p= 3, 5, \dots .  \ee
The most convenient choice of $ \xi_{(z, s)}$, which we adopt in the
following, is
     \be \la{defxis} \xi _{(z,s)} 
     = \sqrt{2} \[M_p^{[3]} \]^{1/3}  \sqrt{ x(z) - x(a_s)}. 
    \ee
Then the moments \re{defmus} at the branch point $a_s$ are given by
 \be
 \la{defmomu}
  \mu_p^{[s]} = \( M_3 ^{[s]} \)^{-p/3} M_{p} ^{[s]} .  \ee

 The left state $\langle \Sigma |$ in the operator representation
 \re{defSigma} is
  \be\la{defVacmb} \langle \Sigma |&=& \langle 0_\tw |\exp\( \hf \sum
   {1\over p\, q}B_{p, q}^{[s,s']} J_p^{[s]} J_q^{[s']} + 
 {1\over \hbar}  \sum {1\over p}\mu_ {p}^{[s]}
    J_{p}^{[s]} \) , \ee
 where $s, s'$ take values $1,\dots, \nb$ and $p, q$ are positive odd
 integers.  By convention $\mu^{[s]} _{p} =0$ if $p\le 1$.

We are looking for a right state of the form
 \be \la{rightstateOm} |\Omega\rangle = \Omega |0_\tw\rangle, \quad
 \Omega = \bigotimes_{s=1}^{\nb} \Omega ^{[s]} , \ee
 where $\Omega^{[s]} $ is the dressing operator associated with the
 $s$-th branch point.  The operator $\Omega^{[s]} $ is determined from
 the conformal Ward identity
 \be\la{confWIFock} \langle\Sigma | L_{2n} ^{[s]} \ \Omega^{[s]}
 |0_\tw ^{[s]} \rangle =0 \qquad (n\ge -1; \ s=1,\dots, \nb), \ee
where $L^{[s]}_{2n}$ are Virasoro operators associated with the expansion of
the twisted stress-energy tensor \re{defTztw}:
\be
T^{[s]}_\tw(z) = \sum _{n\in\IZ} L^{[s]}_{2n} \, \xi_{(z, s)} ^{-2n - 4}.
\ee
The explicit expression for the Virasoro operators is
 \be\la{lmOmLn} L_{2n} ^{[s]} &=& {1\over 4} \sum_{p+q=2n}
 : J_p^{[s]}    J_{q}^{[s]}  : + {1\over 16 }\delta_{n,0}
 \qquad (n\ge -1) .
  \ee
  In order to satisfy  \re{confWIFock} it is sufficient to solve 
   the  operator equation
  \be\la{vViI} e^{{1\over 3 \hbar } J^{[s]}_3} L_{2n}^{[s]} \, \hat
  \Omega^{[s]} \, |0_\tw^{[s]} \rangle = 0 \qquad (n\ge -1, \ s=1,
  \dots, \nb).\ee
Here we retained only the factor $\exp ( {1\over 3\hbar } \mu_3^{[s]}
J^{[s]})$, with $\mu_3^{[s]} = 1$, from the operator deforming the
left vacuum in \re{defVacmb}.  This operator, when commuted to the
right, shifts $J^{[s]}_p \to J^{[s]}_p + \d_{p+ 3, 0}$, which
corresponds to the `minimal' classical solution having a branch point
at $x_s$: $\Phi^{[s]} _\cl\sim \hbar^{-1}  (x-x_s)^{3/2}$.

  Assuming that the  dressing operator $\Omega^{[s]}$ can be expanded as
a formal series in the creation operators $J_{-p}^{[s]} \ (p\ge 1)$,
the solution of \re{vViI} is given by
   \be\la{OmegasE} \Omega ^{[s]}& =& C^{[s]}\ \exp\( \sum _{n\ge 1}
   {(-1)^n\over n!} \sum_{ p_1, \dots, p_n} \hbar^{2g-2+n}\ {w^{(g)}
   _{p_1,\dots , p_n}\over p_1\dots p_n}\ J_{-p_1} ^{[s]}\dots
   J_{-p_n} ^{[s]} \) \!\!  , \ee
where $C^{[s]}$ is a numerical factor depending on the moduli at the
point $a_s$.  The sum goes over odd positive odd integers $p_i$.  The
coefficients $w^{(g)}_{p_1, \dots, p_n}$ are universal rational
numbers and are proportional to the correlation functions in the
Kontsevich model \cite{Kontsevich:1992ti}.  They are nonzero only if
the genus $g$ defined by
\be\la{scc} {1\over 3} \sum _{i=1}^n p_i = 2g-2+n \ee
is a positive integer.  For the sake of self-consistency, we 
derive  in Appendix \ref{appendixA}
 the recurrence equation for $w^{(g)}_{p_1, \dots, p_n}$.
 
The numerical factor $C^{[s]}$ is fixed by the scale invariance.  Upon
a rescaling $x\to \rho x$, the twisted vacuum acquires a factor
$\rho^{1/16}$ and $M^{[s]}_3\to \rho ^{-3/2} M^{[s]}_3$, while the
dimensionless moments $\mu^{[s]}_p$ do not change.  The dressed
twisted vacuum remains scale invariant if
 \be C^{[s]} = (M^{[s]}_3)^{-1/24}.  \ee

\subsection{Summary: a universal formula for the $\hbar$ expansion}

\def\inter{ \text{int}} {\it Partition function}

The free energy $\CF_\hbar =\ln \CZ_\hbar $ is a sum of three terms,
\be \CF_\hbar = \CF_\cl+\CF_\gauss+\CF_\inter , \ee
where $\CF_\cl = \hbar^{-2}\CF^{(0)}$ is the classical action of the gaussian
field on the Riemann surface,
\be \CF_\gauss = \CF^{(1)} =- \ln \det \bar \p - {1\over 24} \sum _{
s=1}^{\nb} \ln M^{[s]}_3 \ee
is the sum of the gaussian fluctuations and the contributions from the
scaling factors associated with the branch points, and $\CF_\inter $
is the interaction part, which vanishes when $\hbar \to 0$.

 Assuming that $\CF_\inter $ is given by an asymptotic series of the
 form
\be \CF_\inter = \sum_{g\ge 2} \hbar ^{2g-2} \CF^{(g)}, \ee
the exponent $\CZ_\inter = e^{\CF_\inter }$ can be evaluated as the
scalar product
  \be\la{FockZ} \CZ_\inter = \langle \Sigma\, | \O\rangle.  \ee
The  states $ \langle \Sigma\, | $ and $ | \O\rangle $ 
belong to the completion of the Fock space
associated with the $\nb$ branch points and defined by the relations
\be [J^{[s]}_p, J^{[s']}_q] &=& p \, \d_{p+q, 0} \d_{s,s'}, \qquad
(s=1,\dots, \nb; \ \ p \in 2\IZ + 1 ) \no
\\
\langle 0_\tw | J^{[s]}_{-p}&=& J^{[s]}_{p}|0_\tw \rangle =0 \qquad
\(s=1,\dots, \nb; \ \ p= 1, 3,5,\dots\).  \ee

    The  left  state in \re{FockZ},
   \be
   \la{defsigma}
    \langle \Sigma\, | &\defeq& \langle 0_\tw |\exp\( {1\over 2}
   \sum_{ s, s'=1}^{\nb} \sum _{p, q \ge 1} {1\over p\, q} B_{p,
   q}^{[s,s']} J_p^{[s]} J_q^{[s']} + {1\over \hbar}\sum_{s=1}^{\nb} \sum_{p\ge 5}
   \mu^{[s]}_p J^{[s]}_p \), \ee
depends on the Bergmann kernel $B$ and the classical field $\Phi_\cl$
through
 \be B_{p,q}^{[s,s']} &=& \Res_{z \to a_s}\Res_{z' \to a_{s'}}
 \xi_{(z,s)}^{p} \xi_{(z',s')}^{q} B(z,z'), \quad p,q= 1, 3, 5, \dots
 ; \no \\
  \mu ^{(s)}_p &=& \Res_{z\to a_s} y(z)d x(z)  \ \xi_{(z,s)} ^{-p},
  \quad p= 5, 7, \dots .  \la{defmusa} \ee

The right  state is  in \re{FockZ},
\be \la{defomega}
 \la{OmegasEb}
 |  \Omega \rangle 
  & \defeq& 
  \prod_{s=1}^{\nb} 
  \exp\( \sum _{n\ge 1}  {(-1)^n\over n!} \sum_{ p_1, \dots, p_n}
   \hbar^{2g-2+n} \
   {w^{(g)} _{p_1,\dots ,  p_n}\over p_1\dots p_n}\
   J_{-p_1} ^{[s]}\dots   J_{-p_n} ^{[s]}   \) \!\! ,  \ee
represents a product of dressing operators, associated with the branch
points.  The coefficients $w^{(g)} _{p_1,\dots , p_n}$ are universal
numbers and are given by the genus-$g$, $n$-loop correlation functions
in the Kontsevich model (see Appendix \ref{appendixA}).

 \bigskip
 
 \noindent
 {\it Correlation functions}
 
 \bigskip
 
 The Fock space representation extends to the 
 correlation functions of the collective field.
 They  are given by
 \be\la{corcp} \< d\Phi_\qu(z_1)\dots d\Phi_\qu(z_n)\>= \<
 d\Phi_\qu(z_1)\dots d\Phi_\qu(z_n)\>_\gauss+ \< d\Phi_\qu(z_1)\dots
 d\Phi_\qu(z_n)\>_\inter \ee
where $\<\ \>_\gauss$ is the result of all possible gaussian
contractions with the Bergman kernel and
 \be\la{corfin} \< d\Phi_\qu (z_1)\dots d\Phi_\qu(z_n)\>_\inter =
 \sum_{s_i =1}^{\nb} \sum _{p_i\ge 1, \netwo} \ B^{[s_1]}_{p_1}(z_1)
 \dots B^{[s_n]}_{p_n}(z_n) \, \langle \Sigma\big|
 J^{[s_1]}_{p_1}\dots J^{[s_n]}_{p_n} \big| \Omega \rangle.  \ee

   \subsection{Diagram technique}

Performing the gaussian contractions we obtain an expression of the
partition function as a series of Feynman graphs made of the following
elements:
   
   \vskip 1cm
   
   Propagator: \be\la{propaa} \begin{array}{c}{\mbox{
   \epsfxsize=3.2truecm \epsfbox{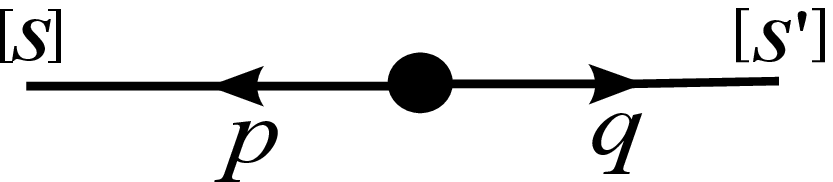} } }\end{array} =
   {B_{p, q}^{[s \, s']}\over p\, q} \, .  \ee

   Tadpole:
  \be\la{tadpolea} \begin{array}{c}{\mbox{
  \epsfxsize=1.2truecm\epsfbox{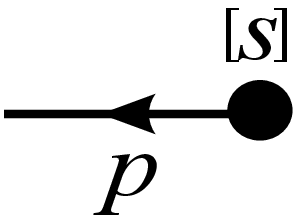} }} \end{array} = -
  {1\over \hbar} \ {\mu_p^{[s]} \over p}.  \ee

    Vertices: \bigskip \be\la{Wnb}
\begin{array}{c}{\mbox{ \epsfxsize=2.7truecm\epsfbox{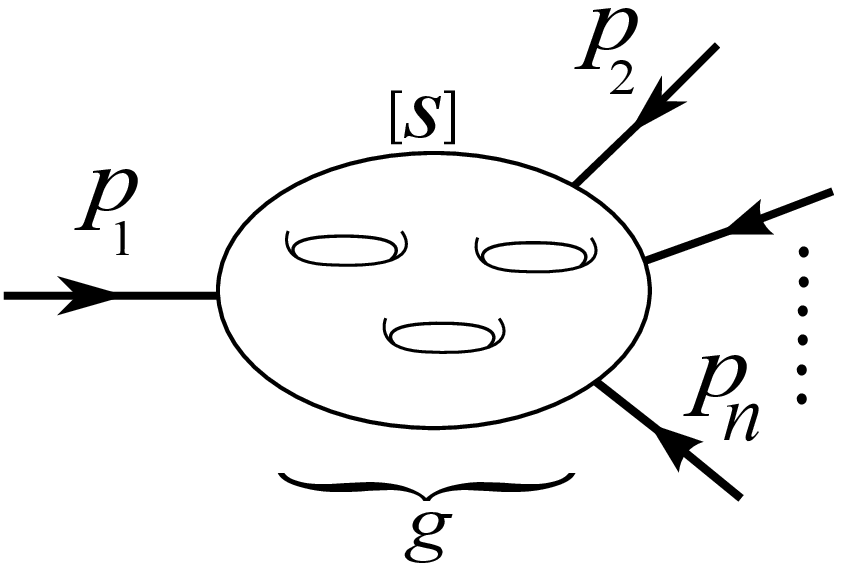}}
}\end{array}= \hbar^{2g-2 +n} \, w^{(g)}_{p\,p_1 \dots p_n } \, , \ee

The tadpoles and the vertices are associated with a given branch point
and carry label $s$, but their weights are given by universal numbers.
The Feynman graphs are made out of these elements by connecting the
open lines respecting the orientation.  The orientation of the lines
is such that vertices connect to tadpoles or propagators, but not
directly with other vertices.  The genus $g$ free energy is a sum of
all connected Feynman graphs of genus $g$.  The factors $(-1)^n$ in
the dressing operator are taken into account by taking the tadpole
with minus sign.
         
The diagram expansion for the correlation functions \re{corcp} is
obtained by adding to the Feynman rules the propagators
	 \be
	   \begin{array}{c}{\mbox{
	   \epsfxsize=1.7truecm\epsfbox{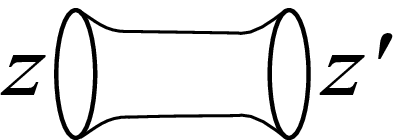}}
}\end{array} = B(z,z') \ee
	  \be\la{Propagaa}
	 \begin{array}{c}{\mbox{
	 \epsfxsize=1.7truecm\epsfbox{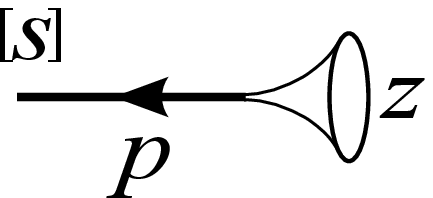}}
}\end{array}= B^{[s]}_p(z) \ee

The Feynman rules \re{propaa}--\re{Wnb} generalize those obtained in
the one-matrix model \cite{Kostov:2009aa} and for the scaling limit of
various matrix models in \cite{Kostov:1991cg, Kostov:1992ie,
Higuchi:1995pv, Kostov:2007xw}.  The structure of the genus expansion
is such that in the case of several branch points the partition
function decomposes to a product of Kontsevich $\tau$-functions,
related by gaussian correlations.  Similar decomposition formulas for
the case of the hermitian matrix model were suggested in
\cite{givental-2000, A.Alexandrov:2009aa, Alexandrov:2008yq,
A.Alexandrov:2007aa}.

It is possible, as is the case for the $O(n)$ matrix model
 \cite{Kostov:1988fy}, that the spectral
curve has a symmetry relating different sheets of the Riemann surface.
If such a symmetry is present, one should identify the twisted vacua
associated with the branch points which belong to the same orbit of
the symmetry.

 \section{Topological recursion  }

\subsection{ Summary}

Here we summarize the definitions and the rules of the method called
topological recursion \cite{Eynard:2004mh, Eynard:2005aa,
Chekhov:2005rr, Chekhov:2006vd, Chekhov:2006rq, Eynard-formal06,
Eynard:2007kz, Eynard:2009zz, borot-2009, Eynard:2008he,
1751-8121-42-29-293001}.

\bigskip

 \noindent 
 {\it Loop  observables}

Given the spectral curve \re{defSC}, we define the $n$-point, genus
$g$, functions $W_n^{(g)}(z_1, \dots, z_n)$ as follows:
\bea\la{Wone} W_1^{(0)}(z) &\equiv& y(z) dx(z) , \cr
W_2^{(0)}(z_1,z_2) &\equiv& B(z_1,z_2) \la{Wtwo} \eea
and  recursively $W_{n+1}^{(g)}(z_0, z_I)\equiv 
W_{n+1}^{(g)}(z_0,z_1,\dots,z_n)$
as
\be\!\!\!  W_{n+1}^{(g)}(z_0,z_I) \!  = \!  - \!  \sum_{s=1}^{\nb}
\Res_{z\to a_s} \!  K(z_0, z) \Big[ \sum _{m, j} \sum_{J\subset I}
W_{j+1}^{(m)}(z,z_J)W_{n-j+1}^{(g-m)}(\tilde{z},z_{I\backslash J}) \!
+\, W_{n+2}^{(g-1)}(z,\tilde{z},z_I\!  ) \!  \Big] .  \ \la{TR} \ee
 The sum inside the brackets goes over $0\le m\le g$, $0\le j\le n$,
 $(m, j)\ne (0,0), (g,n)$, and all possible ways to split the
 variables $z_I=\{ z_1, \dots, z_n\} $ into two sets, $z_J$ and $z_{I
 \backslash J}$.  The r.h.s. is a sum of terms associated with the
 branch points $a_s$ of the Riemann surface $\Sigma$.  Near each
 branch point $\tilde z\ne z$ is defined so that $x(\tilde z) = x(z)$.
 The recursion kernel $K$ is defined in the vicinity of the branch
 point $z=a_s$ as
 \be\la{defK} K(z_0, z) = K(z_0, \tilde{z}) \defeq  {\int
 _{\tilde{z}}^{z} B(z, z_0)\over 2 (y(z)-y(\tilde{z})) dx(z)} , \qquad
 z \text{ close to } a_s .  \ee
The result of the integration does not depend on the choice of the
local coordinate near $a_s$.

\bigskip

\noindent {\it Free energy }
\\
The free energy $\CF$ is defined as the formal (generically
asymptotic) series
\be \CF \equiv \sum_{g=0}^\infty  \hb ^{2-2g} \,  \CF^{(g)}, \ee
whose coefficients are expressed in terms of the loop observables and
the elements of the spectral curve.  $\CF^{(0)}$ and $\CF^{(1)}$ are
respectively the classical action and the gaussian fluctuations of the
bosonic field on the Riemann surface and for $g> 1$: 
\beq \CF^{(g)}
\equiv {1\over 2-2g}\,\sum_s \Res_{z\to a_s}{S(z)  }
W_1^{(g)}(z) \eeq 
where $dS(z) =  y(z) dx(z)$.

\bigskip

\noindent
{\it 
Homogeneity }
\\
Under a rescaling of the one form $ydx$, the free energies turn into
$$
\CF^{(g)}\left[ \lambda \, ydx\right] = \lambda^{2-2g} \CF^{(g)}\left[
ydx\right]
$$
for any $\lambda \in \mathbb{C}$.  

\bigskip

\noindent {\it Symplectic invariance}
\\
For $g >1$, the free energy $\CF^{(g)}$ is invariant under canonical
transformations preserving the Poisson bracket \re{poissbk}.

\subsection{Topological recursion in the vicinity of a branch point}

 The recursion equation \re{TR} involves a sum over the
 branch points of the Riemann surface $\Sigma$.  If we take the point $z_0$ close to one of the branch
 points and take an appropriate scaling limit, the recursion kernel
 will consist of a single term, associated with this branch point.
 The piece of the algebraic curve which survives in the scaling limit
 is a curve with only one simple branch point and genus zero.  This is
 the complex curve for the Kontsevich model.  We will see that the
 solution of the recursion equation for the Kontsevich model plays the
 role of a building block for the topological recursion on an
 arbitrary spectral curve.

\subsubsection{ Solution of the recursion equation for the Kontsevich
model}

 The spectral curve for the Kontsevich integral is a genus 0 spectral
 curve with one simple branch point at $x=x_0$ and singular point at
 $x=\infty$.  It can thus be described by a rational parameterization
 \beq\la{defKonts} \CE : \quad \left\{
\begin{array}{l}
x(z)={1\over 2} z^2 +x_0\cr y(z) = \mu_3 z + \mu_5 z^3 + \mu_7 z^5
+\dots .\cr
\end{array}
\right.  \eeq
The moduli of the spectral curve are the moments $\mu_3, \mu_5, \dots$
and the position $x_0$ of the branch point.\footnote{Of course, the
point $x_0$ can be placed at the origin by a global conformal
transformation.} When we deform the Riemann surface, both the moments
and the position of $x_0$ are varied.
There is only one branch point at $x=x_0$ and the variable $\xi_{(z,
0)} \defeq \sqrt{ 2x-2x_0}=z$ gives a global parametrization of the
curve.

The initial data for the topological recursion are thus the one-form
\beq W_1^{(0)}(z) = y(z) dx(z)=\sum_{p\ge 3, \, \text{odd}}
{\mu_{p}}\, z^{p-1} dz \eeq
(the sum goes over $p$ odd) and the two-form
\beq\la{W02} W_2^{(0)}(z_1,z_2) = {dz_1 dz_2 \over \left(z_1-
z_2\right)^2}.  \eeq

Let us  write down the topological recursion in these terms.
The  recursion equation for the spectral curve \re{defKonts} reads 
\be\la{recureqR}
 \!\! W_{k+1}^{(g)}(z_0,z_K)
=- \Res_{z\to 0}\! 
 K(z_0,z)\!  \Big[ 
 \sum_{m, j, J}  \! W_{j+1}^{(m)}(z,z_J)W_{k-j+1}^{(g-m)}(\tilde{z},z_{K/J})
+ W_{k+2}^{(g-1)}(z,\tilde{z},z_K) \Big]  , \ee
where $\tilde{z} = - z$ and the recursion kernel  \re{defK}  given by 
the series
\be\la{K0gen}
&&K(z_0, z) =  {dz_0 \over 4  y(z) dx(z)} \ {2z\over z^2-z_0^2}
\no \\
&&= - {dz_0\over 2 \, \mu_3 \, z_0^2 z dz} \ \ {1\over 1- {z^2 /z_0^2}} \
\ {1\over 1+ {\displaystyle \sum _{p\ge 3 , \, \text{odd}}}\ (\mu_{p}/
\mu_3) \, z^{p-3} } .  \ee
Since the residue is taken when $z\to 0$ at fixed $z_0$, we have to
expand in the positive powers in $z$.  Apart of the overall power
$1/z$, the expansion is a Taylor series in $z^2$.  If the r.h.s. of
the recursion equation contains only finite number of negative powers
of $z$, which we will see to be the case, then the residue picks only
finite number of terms.

Let us demonstrate how the recursion equation works in the lower
orders.  We will evaluate $W^{(0)}_3$ and $W^{(1)}_1$.  Expanding
\be
B(z, z_1) = {dz\, dz_1 \over z_1^2} (1+ 2 {z\over z_1} + 3 {z^2\over
z_1^2}+\dots) , \ee we find \be W^{(0)} _3 (z_0,z_1,z_3) &=& -
\Res_{z\to 0} K(z_0,z)\, \[B( z, z_1) \, B( -z, z_2)+B( -z, z_1) \, B(
z, z_2)\]\no\\
&=& {1\over \mu_{3}} {d z_0 d z_1 d z_2 \over z_0^2 z_1^2 z_2^2} \\
 W^{(1)} _1 (z_0)\ \ \ &=& - \Res_{z\to 0} K(z_0,z)\, B( z, -z) \no\\
 &=& {1\over 8 \mu_3 z_0^4} -{\mu_5\over 8 \mu_3^2 z_0^2}.  \ee
 It is clear from the form of the kernel that the correlation
 functions with negative Euler characteristics, $2-2g-n< 0$, can be
 written under the form
\beq\hskip -0.3cm \la{Wnseriesb} W_n^{(g)}(z_1,\dots,z_n) = \mu
_3^{2-2g-n-l } \ \sum_{p_i\ge 1 , \, \text{odd}} {\displaystyle
\prod_{i=1}^n} \ { d z_i\over {z_i^{p_i +1} }} \sum_{l \geq 0}{1\over l!}
\sum_{k_j\ge 3} {\displaystyle \prod_{j=1}^l} \(-{ \mu_{k_j} \over k_j
}\) \ \w_{k_1,\dots,k_l|p_1,\dots,p_n}^{(g)}\ \\
, \eeq
where the summation goes over odd integer $p_1, \dots, p_n$ and
$k_1,\dots, k_l$.
The symmetry of the spectral curve $z\to z/\rho, \mu_{2k}\to
\rho^{2k+1} \mu_{2k}$ means that the observables \re{Wnseries} and the
free energy do not change after this rescaling, which gives the
restriction
\be
\sum_{i=1}^l  (k_j-3) + \sum_{i=1}^n (p_i-3)   = 3(2g-2).
\ee
With these notations, the  two  examples considered above give:
\be\la{exom}
\w_{ 1,1,1}^{(0)} = 1 \virg \w_{3|3}^{(1)} = {3\over 8} \virg
\w_{5|1}^{(1)} = {5 \over 8}.
\ee

\subsubsection{ Open-close duality}

The form of the topological recursion obtained in the previous section
involves both ``open'' moduli $z_i$, which we can visualize as
boundaries on some world sheet, as well as the ``closed" moduli $\mu
_p \ (p\ge 3)$.  The two types of moduli are related by the ``loop
insertion operator" \cite{Ambjorn:1992gw}, which depends non-linearly
on the moments $\mu_p$ and the position of the branch point $x_0$.  To
make our derivation self-consistent, we prove here that the
coefficients $\o^{(g)} _{k_J|p_I}$ depend on the two groups of indices
in a symmetric way:
\be\la{opcldu} \w ^{(g)}_{k_1, \dots, k_l | p_{1},\dots, p_m } = \w
^{(g)}_{k_1, \dots, k_k, p_1, \dots p_m}.  \ee
By analogy with the string states we call this property open-close
duality.
 
The open-close duality follows from the study of the variations of the
correlation functions when the initial condition $ydx$ is perturbed.
Consider the variation
\be
ydx(z) \to ydx(z) + \epsilon {B(z,z_*) \over {dz_*}}
\ee
for some $z_*\in{\mathbb C}$ and $\epsilon$ small enough.  This can be
translated into a shift in the moduli $\mu_3, \mu_5,\dots$:
\be
\mu_{p} \to \mu_{p} - \epsilon \ {p }\ z_*^{-p-1}, \quad p=3,5, \dots.
\ee
Therefore the variation represents another background with the same
position of the branch point and the same Bergman
kernel, for which we can write the same loop equations.

 On the other hand, using that by definition of $B= W^{(0)}_2$, it is
 easy to show \cite{Eynard:2007kz} that the correlation functions
 transform as
 $$
W_{n}^{(g)}(z_1,\dots,z_n) \to W_{n}^{(g)}(z_1,\dots,z_n) + \epsilon
{W_{n+1}^{(g)}(z_*,z_1,\dots,z_n) \over dz_*} + \ O(\epsilon^2).
$$
Comparing the two ways to write the linear term in $\epsilon$ in the
r.h.s. of the recurrence equation, we prove \re{opcldu}.  For example,
instead of \re{exom}, we will write
$$
\w _{1,1,1 }^{(0)} = 1 \virg \w _{3,3 }^{(1)} = {3 \over 8} \virg \w
_{5,1 }^{(1)} = {5 \over 8} \ .
$$

\subsubsection{The topological recursion for the open coefficients: a
trivalent theory}

Consider the simplest curve of the class \re{defKonts}, for which
$\mu_3\ne 0$ ans $\mu_{5}=\mu_7 = \dots = 0$.  In this case the
recursion kernel is
\be K(z_0, z) = -{1\over 2 \mu_3\, z\, dz } \ {1\over z_0^2 -z^2}\, .
\ee
Assume that the loop amplitudes have the expansion \be \la{expWm1}
W_n^{(g)}(z_1,\dots,z_n) = \mu_3^{2-2g-n } \sum_{\{p_i\ge 1, \,
\text{odd}\}} \w_{ p_1,\dots,p_n}^{(g)}\ {\displaystyle \prod_{i=1}^n}
\ { d z_i\over {z_i^{p_i+1} }}\, .  \ee
First we notice that if all functions are even in their arguments,
then only the even part of the amplitude $W^{(0)}_2$ (the Bergman
kernel) will contribute to the residue.  This is so if one of the
arguments of $W^{(0)}_2$ is an external variable.  If one of the
arguments, say $z_1$, is external, then we have to expand assuming $
|z|<|z_1|$ and neglect the odd part of the expansion:
\be \la{expW2} W_2^{(0)}(z,z_1) = {dz dz_1 \over (z-z_1)^2} \ \ \to \
\ {1\over z\, z_1} \sum_{p\ge 1}^\infty p \, z^{p}\, z_1^{-p} , \qquad
|z|<|z_1|,\ee
 where the sum goes over the odd $p$.  Thus we can write for
 $W_2^{(0)}$ an expansion of the form \re{expWm1}, but containing both
 positive and negative odd ``momenta" $p$ and define the Bergmann
 kernel in ``momentum space" as
 \be\la{omega0} \w^{(0)} _{p, p_1} = |p| \, \d_{p+p_1, 0}\, \quad (p,
 p_1\in\IZ) .  \ee

When $n=0$ and $g=1$, the first term of the recurrence equation
\re{recureqR} vanishes, while the second term is $W_2^{(0)}(z,-z)=B(z,
-z)$:
\be\la{W20} W_2^{(0)}(z,-z)={1\over 4 z^2}.  \ee

By expanding the topological recursion relation  \re{recureqR} 
and extracting the coefficient of $\mu_3^{2-2g-n}$, one gets a recursion
relation  for the coefficients
$\w _{ p, p_1,\dots, p_n}^{(g)}$ with $2g+n >2$:
 \be\la{recmomspom} \w _{p,p_{_I}}^{(g)} &=& {1 \over 2 } \sum_{q+k =
 p-3} \ \(\ \sum_{h=0}^{g} \ \sum_{J \subset I} \ \w _{q, \,
 p_{_J}}^{(h)} \ \w _{ k, \, p_{_{I\backslash J}}}^{(g-h)} + \w _{ q ,
 k , \, p_{_I}}^{(g-1)} \) .  \ee
 Here we used the abbreviation $ p_I=\{p_1,\dots,p_{n}\} $ for the set
 of indices $I= \left\{1,\dots, n\right\} $ and the sum goes over all
 possible ways to split the set $I$ into two non-overlapping subsets
 $J$ and $I\backslash J$.  The solution of equation \re{recmomspom}
 can be expressed as a sum of connected Feynman-like diagrams with
 trivalent vertices.  The vertex whose three legs are labelled by
 momenta $p, q, k$ pointing inwards imposes the restriction $
 p+q+k=3$.  All internal vertices and propagators have weight 1 except
 the external lines, which have weight $|p|$.  It will be convenient
 to indicate only the absolute value of the momenta and indicate the
 signs by arrows on the propagators.  Then, representing the
 amplitudes $\w^{(g)} _{ p_1,\dots, p_n} $ as blobs with $g$ handles
 and $n$ legs,
\bigskip \be\la{wnb} \w^{(g)} _{ p_1,\dots, p_n} =
\begin{array}{c}{\mbox{ \epsfxsize=2.2truecm\epsfbox{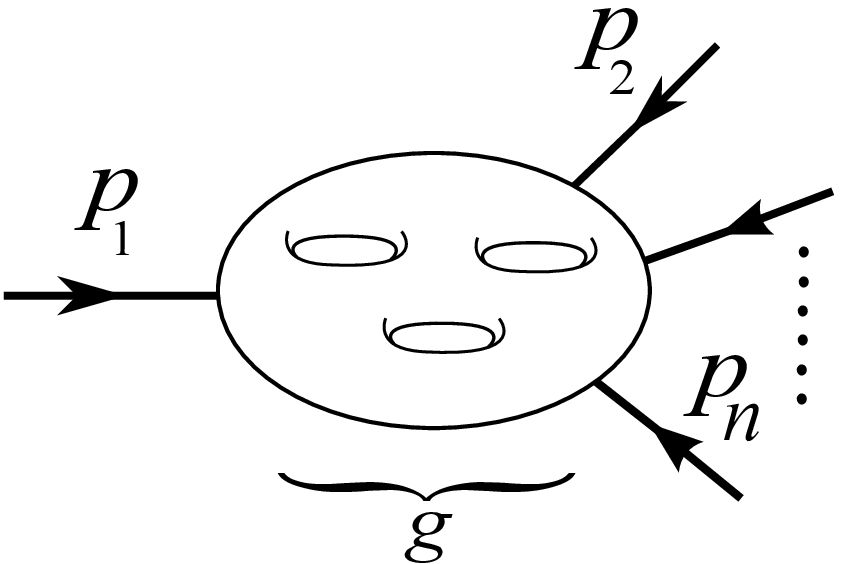}}
}\end{array}\, , \ee
\bigskip
the recurrence equation takes the form 
\bigskip 
\be\la{recureqc}
\begin{array}{c}{ \epsfxsize=10truecm\epsfbox{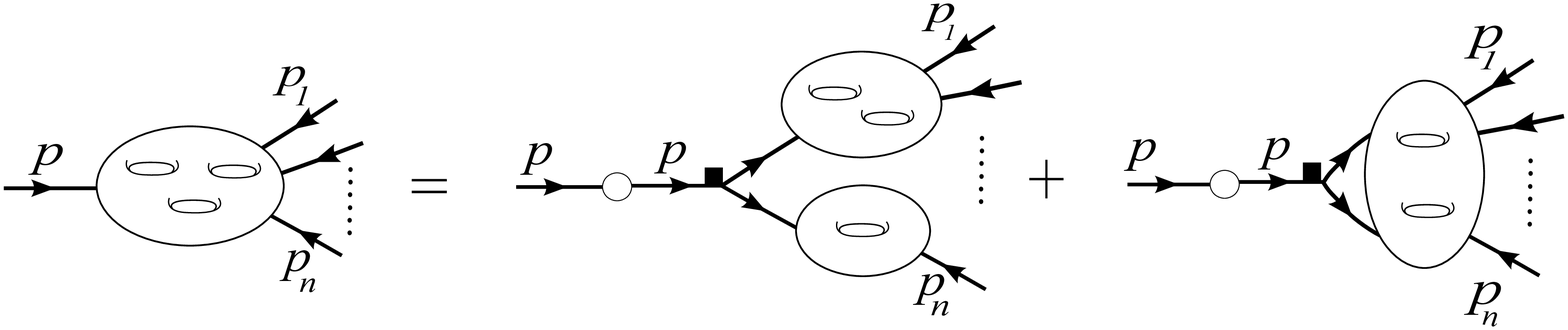}
}\end{array}\, \ .  \ee
\bigskip 
The vertex (with the given orientation of the lines) is
\be \begin{array}{c}{\mbox{ \epsfxsize=1.5truecm\epsfbox{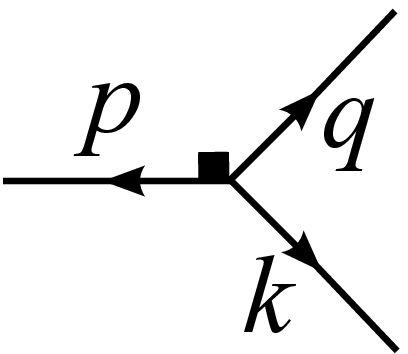} }
}\end{array}= {\d_{p+q+k+ 3,0} \over |p|}\ .  \ee
The black square indicates the momentum whose absolute value is put in
the denominator.  The propagator is the genus zero two-point function
\re{omega0}
\be\la{bareprop} \w^{(0)}_{p, q}\ \equiv \ \
\epsfxsize=2.2truecm\epsfbox{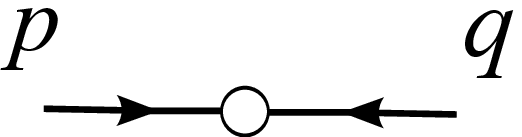}= |p|\, \d_{p+q, 0} =
\epsfxsize=2.0truecm\epsfbox{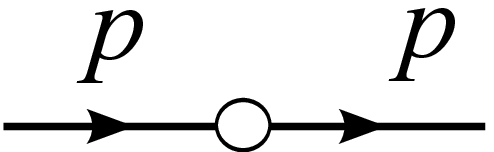} . \ee

The initial conditions are given by the genus zero two-point function
\re{bareprop} and the genus one tadpole, which is computed by
\re{W20}:
\be\la{baretadpole} \w^{(1)}_p \equiv
   \begin{array}{c}{\mbox{ \epsfxsize=1.6truecm\epsfbox{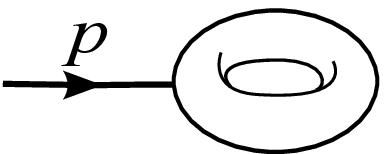} }}
   \end{array}
=
  \begin{array}{c}{\mbox{ \epsfxsize=2.7truecm\epsfbox{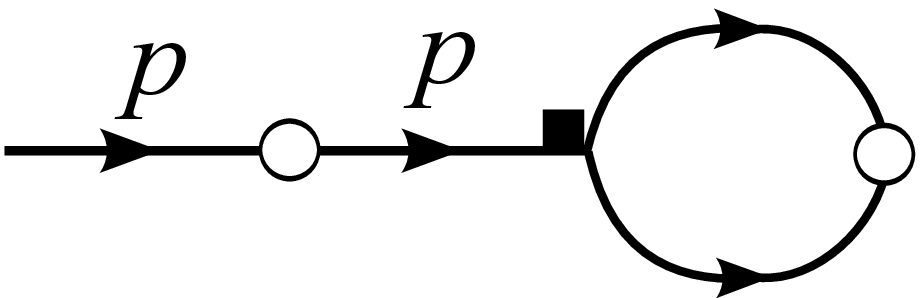} }}
  \end{array}
= { 1\over 8} \, \d_{p, 3} \, .  \ee

{\it Examples:}

 \be \w_{1,1,1}^{(0)} &\equiv&
  \begin{array}{c}{\mbox{ \epsfxsize=4.0truecm\epsfbox{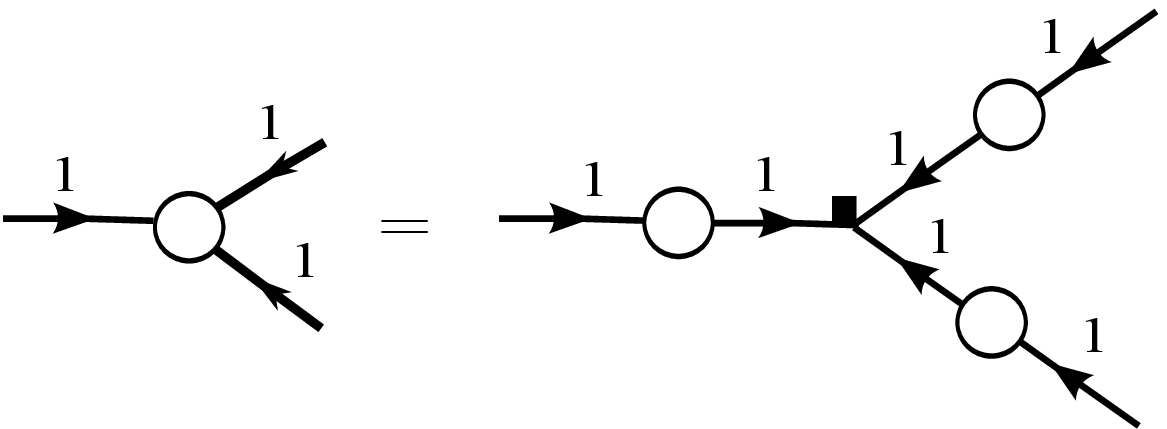} }}
  \end{array}
 = 1, \no\\
 \w^{(0)}_{3,1,1,1} &\equiv&
  \begin{array}{c}{\mbox{ \epsfxsize=4.7truecm\epsfbox{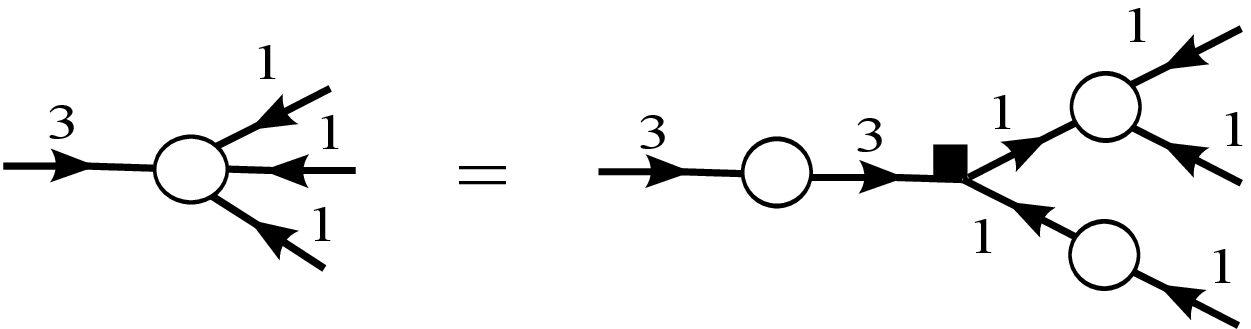} }}
  \end{array}
= 3 \cdot {1\over 3} \cdot \w_{1,1,1}^{(0)} = 1, \no \ee \be
\w^{(0)}_{3,3} &\equiv& \begin{array}{c}{\mbox{
\epsfxsize=4.2truecm\epsfbox{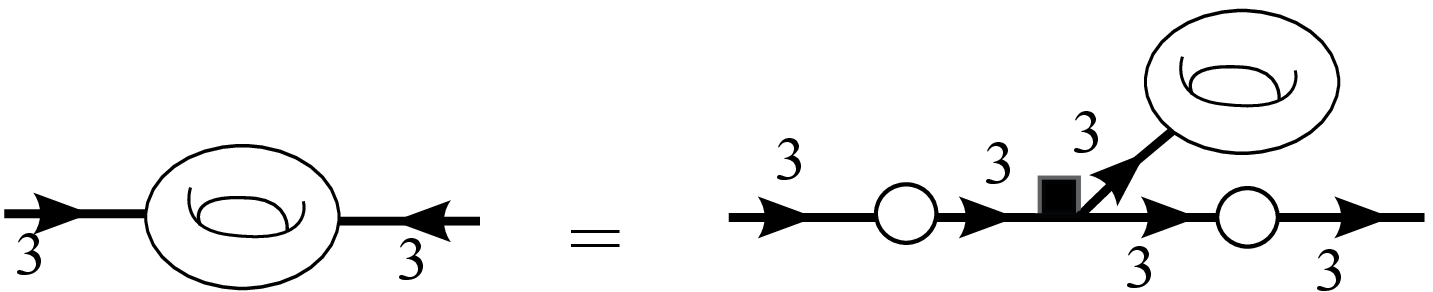} }} \end{array} = 3 \cdot
{1\over 3} \cdot \w_{3}^{(1)} \cdot 3 ={3\over 8}, \no\\
    \w^{(0)}_{1,5} &\equiv& \begin{array}{c}{\mbox{
    \epsfxsize=4.6truecm\epsfbox{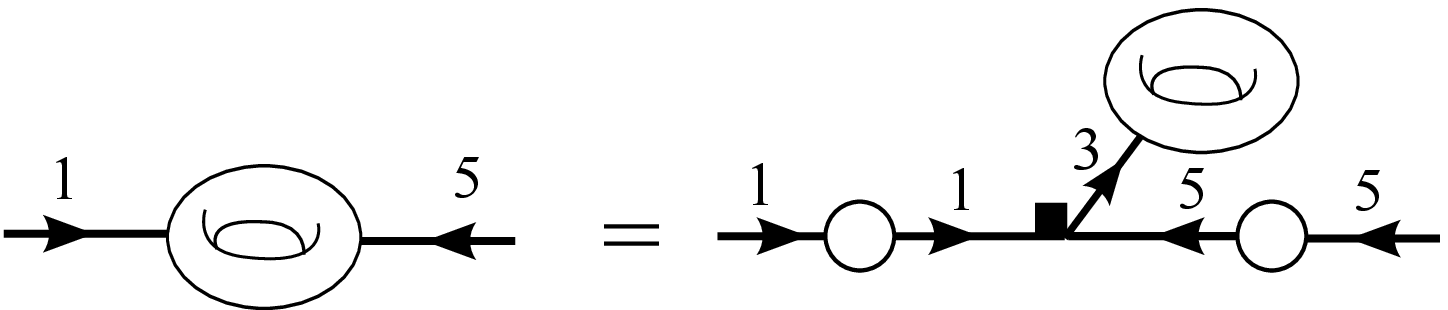} }} \end{array} =
    \w_{3}^{(1)} \cdot 5 = {5\over 8} \no\\
  \w^{(0)}_{5,1} &\equiv& \begin{array}{c}{\mbox{
  \epsfxsize=7.0truecm\epsfbox{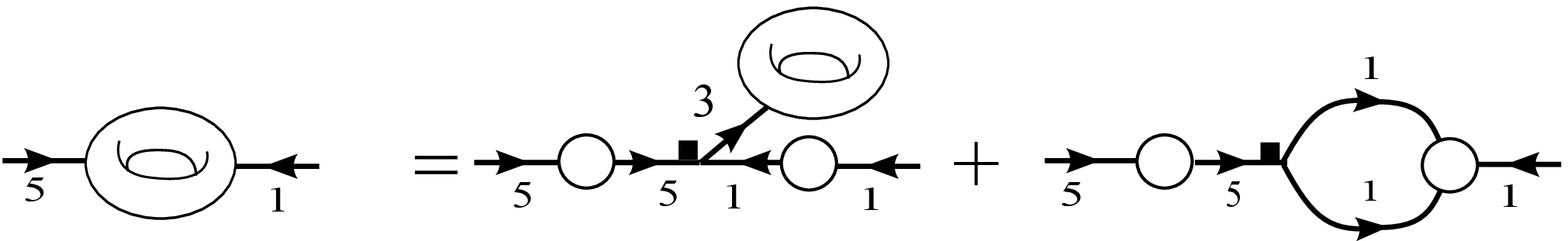} }} \end{array} \no\\ &=& 5
  \cdot {1\over 5} \cdot ( \w_{3}^{(1)} + {1\over 2} \cdot
  w^{(0)}_{1,1,1}) = {1\over 8} + {1\over 2} = {5\over 8} \no\ee \be
  \w_ 9^{(2)} &\equiv&
  \begin{array}{c}{\mbox{ \epsfxsize=10.8truecm\epsfbox{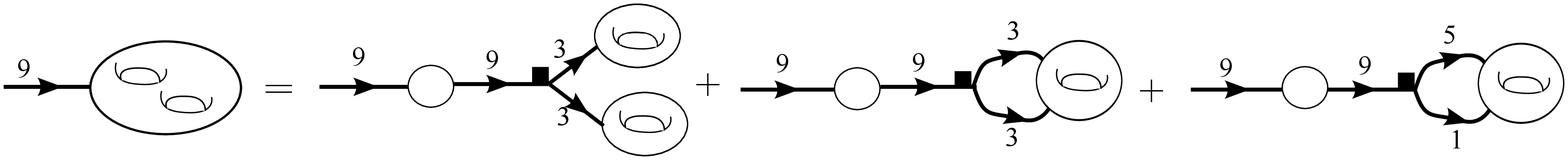} }}
  \end{array}
\no\\
&=& 9 \cdot {1\over 9} \cdot \( {1\over 2} \left(\w_ 3^{(1)}\right)^2+
{1\over 2} \w_ {3,3}^{(1)} + \w_{5,1}^{(1)} \) = {1\over 2} \( {1\over
8^2} + {3\over 8} + 2 {5 \over 8}\) = {105 \over 128}.  \no \ee

 The recurrence equation resembles the mean field equation for a
 theory with cubic potential.  This analogy can be made exact if one
 thinks of the string field theory Hamiltonian, which describes the
 elementary processes of splitting and joining of strings.  Therefore
 the solution is not a series of Feynman diagrams, but have a
 tree-like structure, even though they have loops.  This means that
 any diagram has a root (the starting point of the iteration) and the
 vertices are labeled by their height (the distance from the root of
 the tree).  Because of this structure, the number of graph grows much
 slower that the number of Feynman graphs as the number of loops and
 external legs increases.  These recurrence equations are of course
 equivalent to those obtained in topological gravity
 \cite{dijkgraaf348loop, Dijkgraaf:1990ps, Dijkgraaf:1990nc}.

\subsubsection{Topological recursion for generic values of the moduli}

Once we know the coefficients $w^{(g)}_{p_1,\dots, p_n}$, from the
general form \re{Wnseriesb} of the solution and the open-closed duality
it follows that we can write down the free energy and the correlators
for any background, characterized by the moments $\mu_5,\mu_7, \dots$
by allowing lines with tadpoles.  Consider the generating function
\be\la{frenerg} F^{(g)} [\{t_p\}]=\sum _{n\ge 1} {1\over n!} \mu_3
^{2-2g-n} \sum_{p_1,\dots, p_n} w^{(g)}_{p_1,\dots, p_n} \
{t_{p_1}\over p_1} \dots {t_{p_n}\over p_n}.  \ee
The loop amplitudes \re{Wnseries} are obtained by applying the ``loop
insertion operator''
\be \hat w(z) = \sum_{ p\ge 1} {1\over z^{p+1} }p {\p \over \p t_p}
\ee
to the free energy $F^{(g)} [\{t_p\}]$:
 \be\la{WnF} W_n^{(g)} (z_1,\dots, z_n) &=& dz_1\dots dz_n \hat w(z_1)
 \dots \hat w(z_n) F^{(g)} [\{t_p -\mu_p\}]\Big|_{t_p=0}\no \\
&=& \sum_{p_i\ge 1} W^{(g)}_{p_1,\dots, p_n}\ {dz_1\over
z_1^{p_1+1}}\dots {dz_n\over z_n^{p_n+1} }.  \ee The new coefficients
are graphically represented as follows: \be W^{(g)}_{p_1,\dots, p_n}\
=
   \begin{array}{c}{\mbox{
   \epsfxsize=2.5truecm\epsfbox{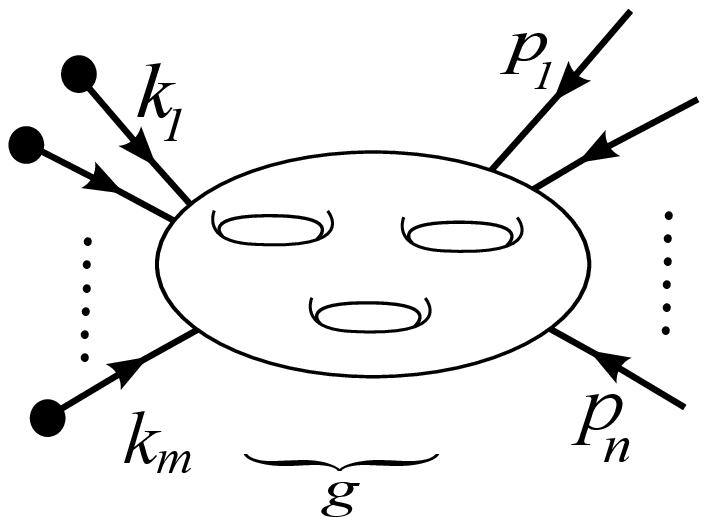}}}\end{array}
\ee
where the black blobs represent the moments $\mu_p$:
  \be\la{tadpole}
  \begin{array}{c}{\mbox{ \epsfxsize=1.3truecm\epsfbox{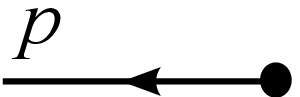} }}
  \end{array}
 =- {\mu_p\over p}\   \qquad (p=5,7, \dots).
  \ee

The above result can be obtained directly from the recursion equation
with the general recursion kernel \re{K0gen}.  By expanding the last
factor in \re{K0gen} we obtain the recurrence equation \re{recureqc}
with a modified vertex, which can be graphically represented as a sum
of all possible insertions:
\bigskip
\be\la{tadpoles}
\begin{array}{c}{
\epsfxsize=4truecm\epsfbox{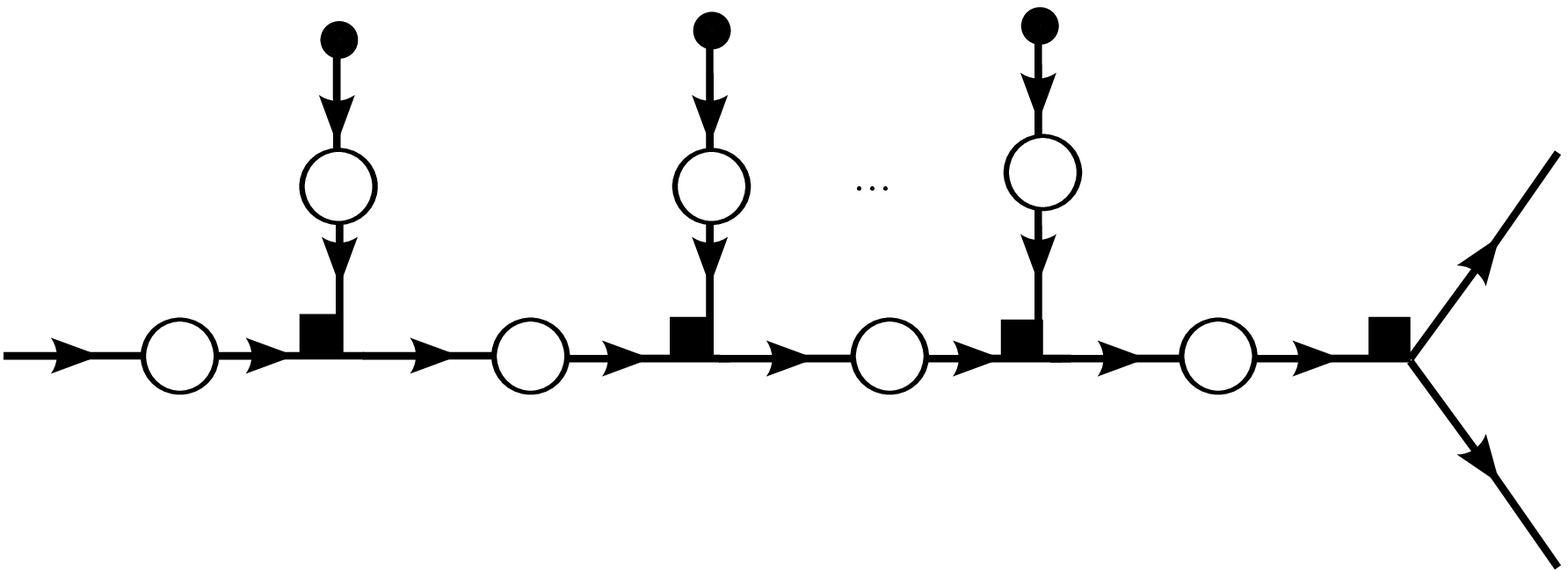}}\end{array}\ .  \ee
\bigskip 
 The same effect will be obtained if we consider each of the
 insertions in the dressed vertex \re{tadpoles} as the result of a
 recursion with the original vertex, with one of the amplitudes given
 by the tadpole
   \be
    \begin{array}{c}{\mbox{ \epsfxsize=1.3truecm\epsfbox{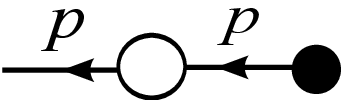}
    }} \end{array}
 = - \mu_p  \ee

In particular the genus $g$ free energy for an arbitrary background
$\left\{\mu_p\right\}$ is 
\be\la{frenergb} \CF^{(g)} =\sum _{n\ge 1}
{1\over n!} \mu_3 ^{2-2g-n} \sum_{p_1,\dots, p_n} w^{(g)}_{p_1,\dots,
p_n} \ {- \mu_{p_1}\over p_1} \dots {- \mu_{p_n} \over p_n}.  \ee

\subsection{Topological recursion for an arbitrary spectral curve}

Now let us consider an arbitrary spectral curve with $\nb$ branch
points $\left\{a_s\right\}_{i=1,\dots,\nb}$ as defined in section 2.
We can cover the Riemann surface by an atlas of chartes associated
with the local coordinates $\xi_{(z,s)}, \ s=1, \dots, \nb$.  The
coordinate $\xi_{(z,s)}$ parametrizes the vicinity of the branch point
$z=a_s$.

In order to write down the topological recursion \re{TR} in a closed
form, we need to expand the loop amplitudes in the local coordinates.
Assume that the points $z_1,\dots, z_n$ are close respectively to
$a_{s_1}, \dots, a_{s_n}$.  Then, assuming that $2g-2 + n>0$, the
$n$-point amplitude is expanded as a sum over odd integer momenta
$p_1,\dots, p_n$:
\be \la{Wnseries} W_n^{(g)}(z_1,\dots,z_n) \Big|_{z_i \to a_{s_i}} = \
\sum_{p_i\ge 1} { W^{(g)}}_{p_1 \dots p_n }^{[s_1 \dots s_n] } \ { d
\xi_{(z_1,s_1)}\over { \xi_{(z_1,s_1)}^{p_1+1}} } \dots { d
\xi_{(z_n,s_n)}\over { \xi_{(z_n,s_n)}^{p_n+1}} } .  \ee
We will represent graphically the amplitudes $ { W^{(g)}}_{p\,p_1
\dots p_n }^{[s \, s_1 \dots s_n] }$ as
\be\la{Wnc} { W^{(g)}}_{p\,p_1 \dots p_n }^{[s \, s_1 \dots s_n] } =
\begin{array}{c}{\mbox{ \epsfxsize=2.7truecm\epsfbox{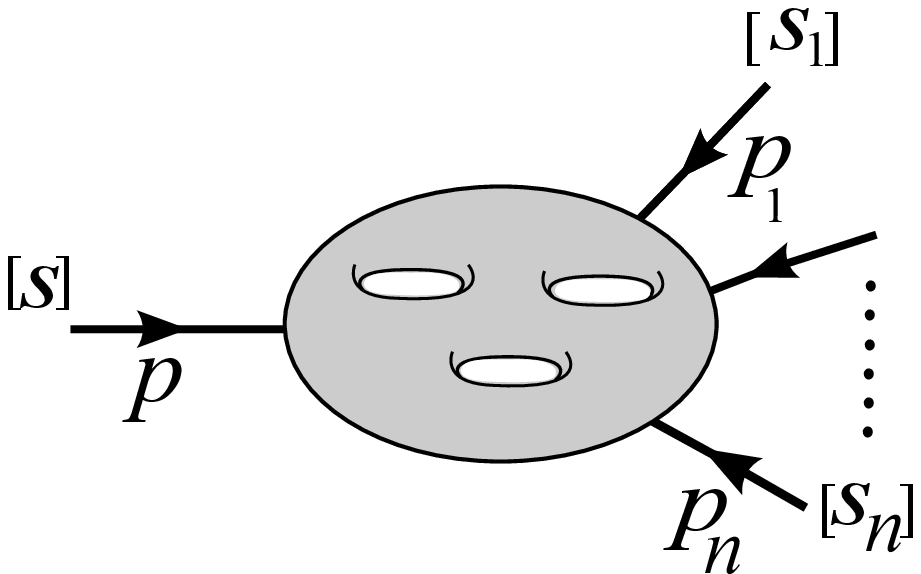}}
}\end{array}\, .  \ee
The momenta $p_i$ are associated with the oriented external legs and
the labels $[s_i]$ are associated with the extremities of the legs.

Consider first a background with $\mu_p^{[s]} = \delta_{p,3}$.  Then
the recurrence equation for the amplitudes is
 \bigskip \be\la{recureqd} \epsfxsize=13truecm\epsfbox{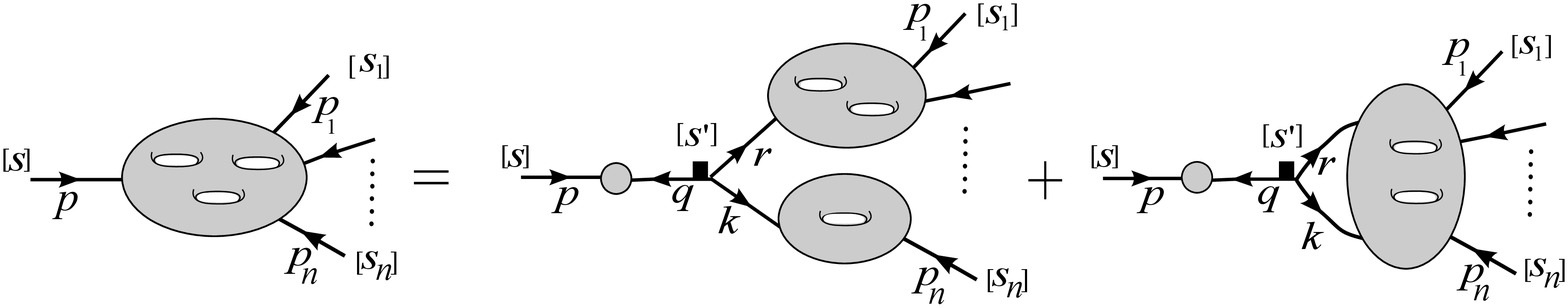}\
 .  \ee
where in the right hand side one sums over $s' = 1, \dots, \nb$ and
all odd integers $q, k$ and $ r$.  The propagator and the vertex are
defined as
\be \begin{array}{c}{\mbox{
\epsfxsize=2.7truecm\epsfbox{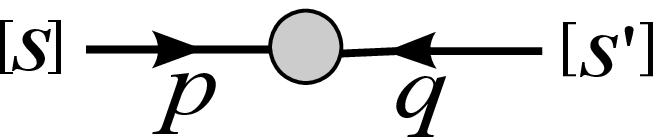} } }\end{array} \equiv
B_{-p,-q}^{[s,s']}&=& \Res_{z\to a_s} \Res_{z\to a_{s'}} { B(z,z')
\over \xi_{(z,s)} ^{p} \xi_{(z', s')} ^{q}} \ , \ee
\be \begin{array}{c}{\mbox{ \epsfxsize=1.5truecm\epsfbox{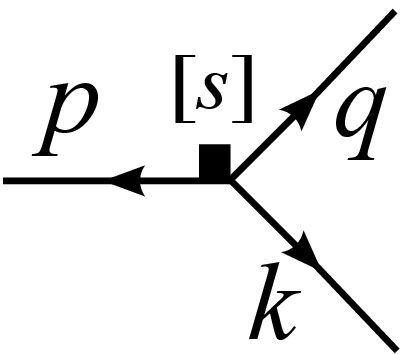}
} }\end{array}= {\d_{p+q+k+ 3, 0} \over |p|}\ .  \ee The propagator
consists of two terms,
 \be \la{blwh} \begin{array}{c}{\mbox{
 \epsfxsize=7.6truecm\epsfbox{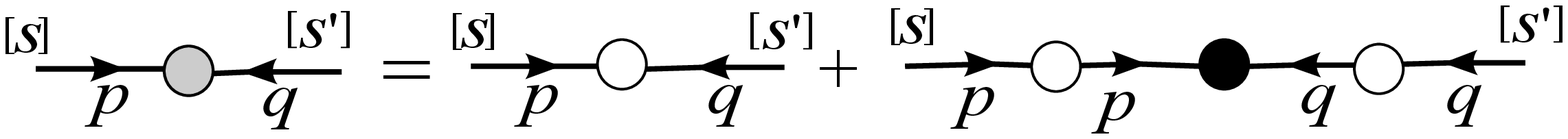} } } \end{array} .
 \ee
 The first term is the diagonal part \re{bareprop}, which conserves
 the momentum:
\be\la{whitepr}
 \begin{array}{c}{\mbox{
 \epsfxsize=5.3truecm\epsfbox{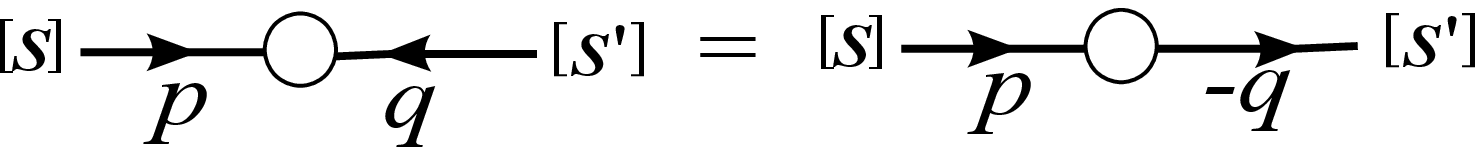} } } \end{array}
 = \d_{p+ q, 0}\, \d_{s, s'} \ |p| .  \ee
 The second term is nonzero when $p$ and $q$ are positive, which is
 illustrated by changing the orientation of the second line,
\be\la{blackpr}
 \begin{array}{c}{\mbox{
 \epsfxsize=5.3truecm\epsfbox{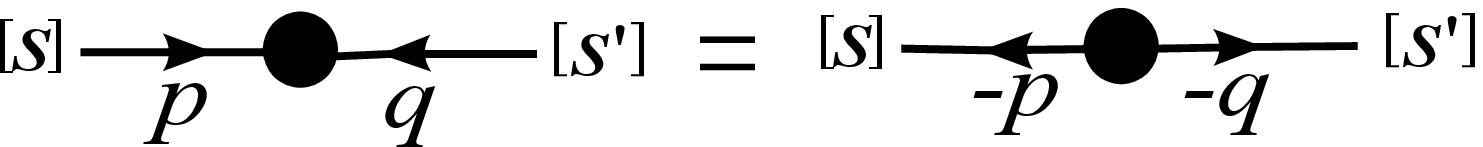} } } \end{array}
 = {B_{-p,-q}^{[s,s']}\over p\ q} .  \ee
We used the same graphical notations as in the previous chapter; the
white blobs are added to compensate the factors $1/p$ and $1/q$ in the
definition of the black blob.
   
Finally, the genus one tadpole, which is computed by \re{W20} does not
depend on $s$:
\be\la{baretadpoleb} W^{(1)}_p \equiv
   \begin{array}{c}{\mbox{ \epsfxsize=1.4truecm\epsfbox{g1n1.eps} }}
   \end{array}
=
  \begin{array}{c}{\mbox{ \epsfxsize=2.4truecm\epsfbox{1loop2.eps} }}
  \end{array}
= { 1\over 8} \, \d_{p, 3} \, .  \ee

\subsubsection{Loop insertion operator}

For a generic background $\{\mu_i^{[s]}\}$, the expansion of the
correlation functions near a set of branch points $a_{s_1},\dots,
a_{s_n}$ reads
 \be W_n^{(g)}(z_1,\dots,z_n) \Big|_{z_i \to a_{s_i}} =
\ \sum_{p_i\ge 1} \sum_{m\geq 0} \sum_{k_j \geq 3 , \, \text{odd}} 
{1\over m!}{
W^{(g)}}_{p_1 \dots p_n| k_1 \dots k_m }^{[s_1 \dots s_n] [s'_1 \dots
s'_m]} \ \prod_{i = 1}^n { d \xi_{(z_i,s_i)}\over {
\xi_{(z_i,s_i)}^{p_i+1}} } \prod_{j = 1}^m \left(- {\mu_{k_j}^{[s_j']}
\over k_j}\right) .  \ee
As in the preceding case, one can add one variable in the correlation
functions by using the loop insertion operator consisting in the shift
of the differential $ydx$ by the Bergman kernel:
\be ydx(z) \to ydx(z) + \epsilon {B(z,z_*) \over dz_*}.  \ee
The only difference with the preceding section is that this operator
acts as a shift on the times around all the branch points at once.
Indeed, it acts as
\be \mu_k^{[s]} \to \mu_k^{[s]} - \epsilon {k \over
\xi_{(z_*,s)}^{k+1}} \qquad \hbox{for all} \qquad s=1,\dots,n_B .  \ee
Applying the loop insertion operator to the genus $g$ free energy,
with the insertion point close to a branch point $a_s$ gives the open
closed duality
\be { W^{(g)}}_{p_1 \dots p_n| k \, k_1 \dots k_m }^{[s_1 \dots s_n]
[s \, s'_1 \dots s'_m]} = { W^{(g)}}_{k \, p_1 \dots p_n| k_1 \dots
k_m }^{[s \, s_1 \dots s_n] [s'_1 \dots s'_m]}\, .  \ee

\subsubsection{Topological recursion for generic background}

The open closed duality allows us to get the expression of the free
energy in an arbitrary background characterized by the moments
$\{\mu_i^{[s]}\}$ at the branch points $a_s, s=1,\dots, \nb$ , once we
know the correlation functions of the trivial background $\mu_p^{[s]}
= \delta_{p,3}$.  Indeed, the free energies
\beq \CF^{(g)} =\sum_{n \geq 0}{1\over n!} \sum_{p_i \geq 3 , \, \text{odd}} \
\sum_{s_i = 1}^{n_B} {W^{(g)}}_{p_1 \dots p_n}^{[s_1 \dots s_n]}
\prod_{i = 1}^n \left(- {\mu_{p_i}^{[s_i]} \over p_i}\right) .  \eeq
 are obtained by attaching tadpols
  \be\la{tadpoleb}
  \begin{array}{c}{\mbox{ \epsfxsize=1.1truecm\epsfbox{Tadpolegen.eps}
  }} \end{array}
 =- {\mu_p^{[s]} \over p}\ \qquad (p=5,7, \dots).  \ee to the external
 legs.

 \bigskip

\section{Relation between the two approaches }

 \noindent {\it Observables:}

 The symplectic invariants defined by \re{Wone}, \re{Wtwo} and \re{TR}
 are related to the correlation functions of the CFT on the Riemann
 surface \re{corfin} by
 \be \la{loopvariables} \< d\Phi(z_1)\dots d\Phi(z_n)\>
 _{\text{connected}} = \sum_{2g+n>2}  \hbar^{2g -2+n}\ W^{(g)} (z_1,\dots,
 z_n).  \ee
 Furthermore, using \re{deftwcor}, the recursion kernel  can be  written 
 as 
 \be
 K(z_0, z) = {1\over 2} 
{  \langle d\Phi_\qu (z_0) \Phi_\qu ^{[s]}(z) \rangle_{_\Sigma}
\over  
   d \Phi_\cl ^{[s]}(z) }.
  \ee

  \bigskip

 \noindent {\it Recursion equation and conformal Ward identity:}
 
The recursion equation represents an integrated form of the Virasoro
constraints.  To see that, we write the recursion equation
\re{recureqR} as the operator identity
 \be\la{recurOp} 2\, d\Phi_\qu (z_0)+\sum_{s=1}^{\nb} \Res_{z\to a_s} \(
 \langle d\Phi_\qu (z_0) \Phi_\qu ^{[s]}(z) \rangle_{_\Sigma} \ {d\Phi
 _\qu (z ) d\Phi _\qu (\tilde z )\over  d \Phi_\cl ^{[s]}(z) }\) =0.
 \la{TRfield} \ee
Inserting this expression in the expectation value \re{loopvariables}
one indeed obtains \re{recureqR}.

 For $z_0$ close to the branch point $a_s$, the two-point function has
 Laurent expansion
 \be \langle d\Phi_\qu (z_0) \Phi_\qu ^{[s]}(z) \rangle_{_\Sigma} & =&
 d \xi_{(z_0,s) } \sum_{p\ge 1,\netwo } \xi_{(z_0,s)} ^{-p-1}
 \xi_{(z,s)} ^p +\text{regular} .  \ee
 Therefore the residue  in \re{recurOp} is a projection to the 
  singular at $z_0=a_s$ part of the integrand. 
  The recursion formula  \re{recurOp} implies that
 \be \la{TRbp}
  2\, d\Phi_\qu (z)     +  {d\Phi _\qu (z ) d\Phi _\qu (\tilde z )\over 
  d \Phi_\cl ^{[s]}(z) }  
  \approx 
  { [d\Phi  ^{[s]}  (z)  ]^2  \over 
   d \Phi_\cl ^{[s]}(z) }   \approx 0,
 \ee
where $d \Phi^{[s]}=d \Phi_\cl ^{[s]}+d \Phi _\qu^{[s]}$ and $\approx$
means `equal up to a regular at $z=a_s$ function'.  Since $d\Phi
_\cl^{[s]}(z) \sim \xi_{(z,s)}^2 d \xi_{(z,s)} $ when $z\to a_s$,
\re{TRbp} is equivalent to the conformal Ward identity \re{confWI}.

 \bigskip

 \noindent
 {\it   Graph expansions:}

The trivalent graphs that appear in the recursion procedure and the
Feynman graphs of the CFT approach must lead to the same result.
Moreover, the Feynman rules of the CFT approach can be obtained by
partial resummation of the trivalent graphs of the recursion
procedure.

To see that, we first notice that, depending on the incoming and the
outgoing momenta, only one of the two terms in the propagator \re{blwh}
can be non-zero.  Any trivalent graph can be split into connected
subgraphs in such a way that the vertices belonging to the same
subgraph can be connected only by Kontsevich (white) propagators,
while the vertices belonging to two different subgraphs can be
connected only by black propagators.  Each such subgraph is associated
with given branch point $a_s$ and contain only Kontsevich propagators
\re{whitepr}.  We can reorganize the sum over trivalent graphs by
first summing up the contributions of the connected subgraphs
associated with the branch points.  We represent the sum of the
contributions of all subgraphs with given values $p_1, \dots , p_n$ of
the external momenta by an effective vertex with $n$ legs and genus
$g$ given by \re{scc}.  For example,
 \be\la{exampleabc}
  \begin{array}{c}{\mbox{ \epsfxsize=5.5truecm\epsfbox{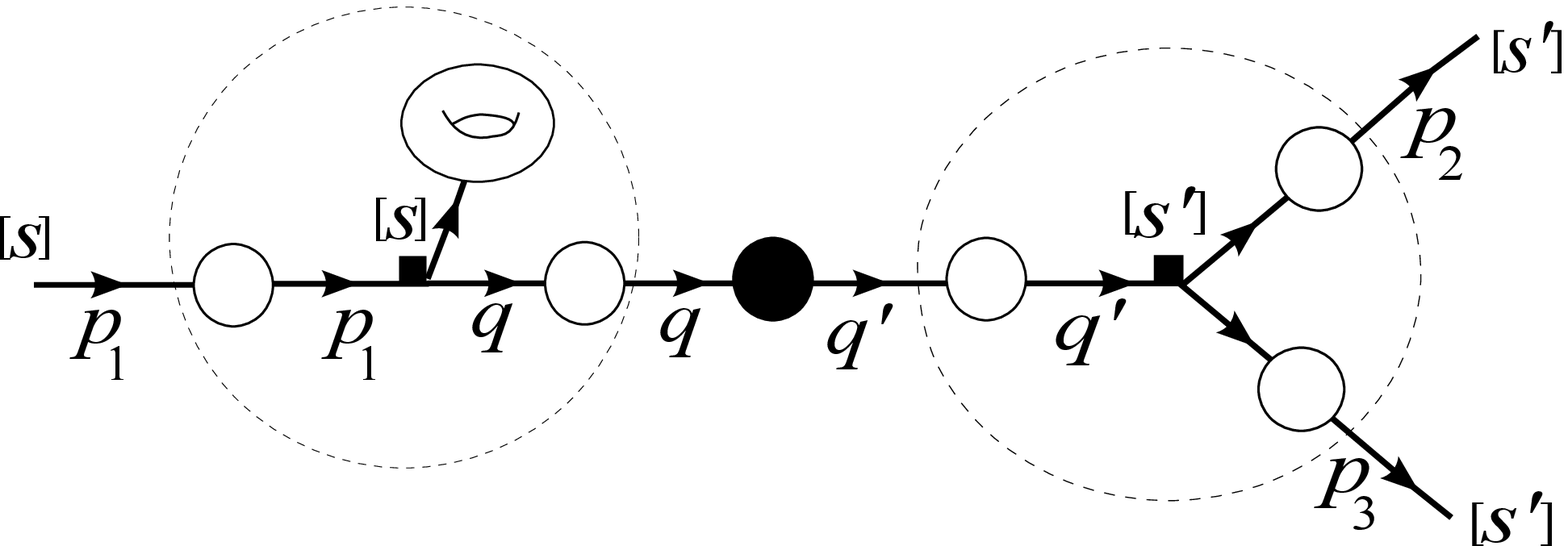}
  }} \end{array}
\!\!\!  &+& \!\!\!  \begin{array}{c}{\mbox{
\epsfxsize6.9truecm\epsfbox{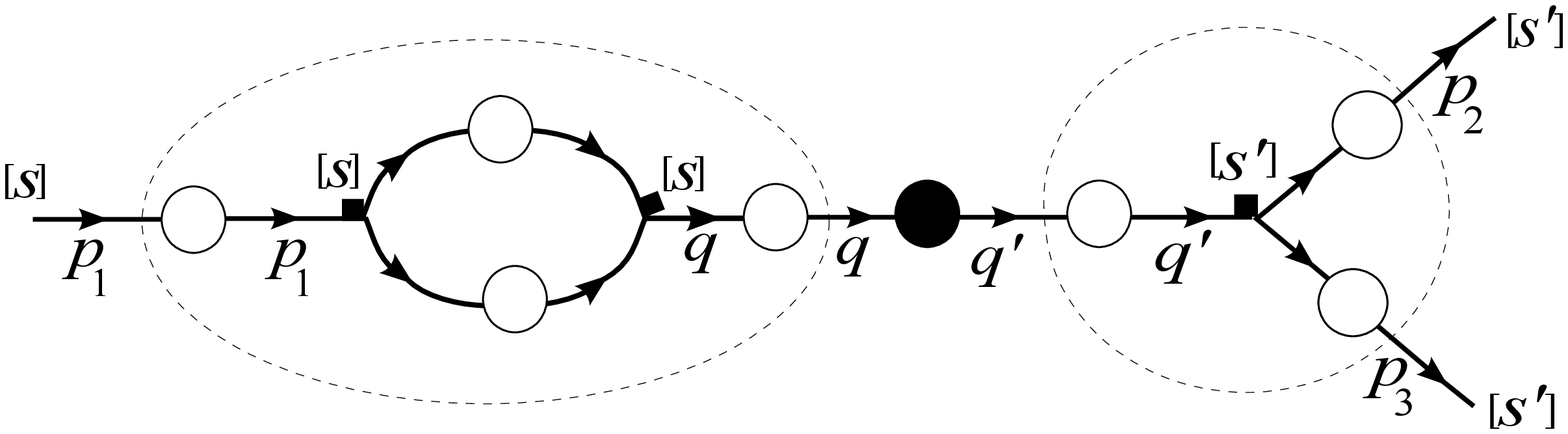} }} \end{array} \no \\
 &=& \begin{array}{c}{\mbox{
 \epsfxsize=4.1truecm\epsfbox{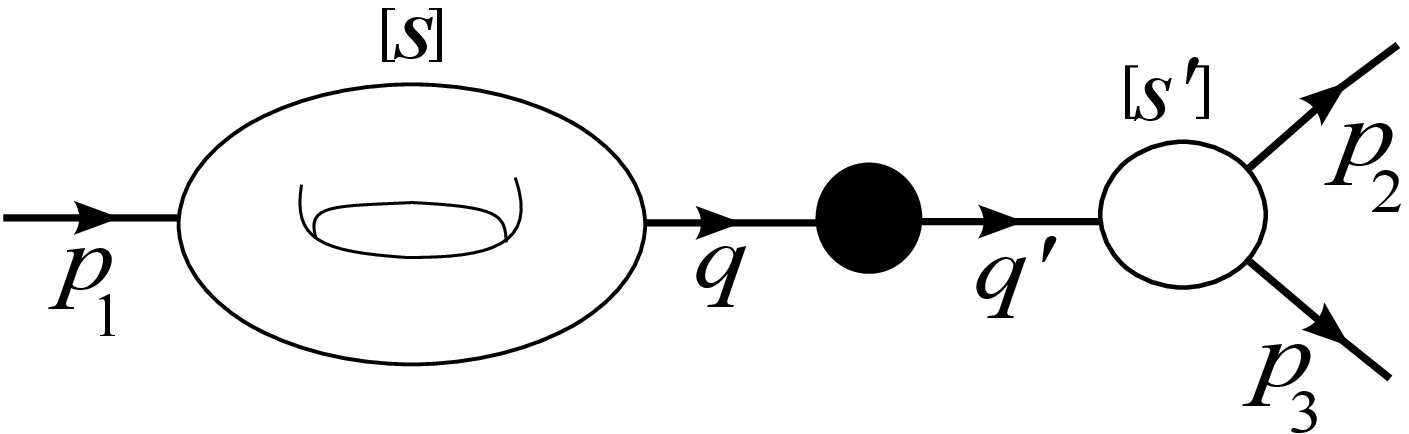} }} \end{array} \!\!\!  .
 \ee

The partial resummation recasts the trivalent theory into an effective
theory whose vertices are universal numbers, the amplitudes
$w_{n_1,\dots,n_k}^{(g)}$ of the Kontsevich model, and whose
propagator and tadpole are given respectively by \re{propaa} and
\re{tadpolea} with $\hbar=1$.

\section{Conclusion}

In this paper we reviewed and compared two approaches to
quasiclassical expansion associated with a spectral curve:
the CFT approach and the topological
recursion (TR).  The two approaches are formulated in their most
general form.  The only input is the spectral curve associated with
the classical solution.  Assuming that the collective field theory is
conformal invariant, we derived a general formula for the genus
expansion of the free energy and the current correlation functions.
 
We demonstrated that the diagram techniques obtained within the two
approaches, although seemingly very different, lead to the same result
for the genus expansion.  More concretely, we showed that the CFT
graph expansion is obtained as a partial resummation of the graph
expansion in the TR.

 The result of this paper proves that any theory solved by the
 topological recursion has a CFT dual living on the corresponding
 spectral curve.  On the other hand, it provides a new way to prove
 that a system can be solved by the topological recursion.  Namely, if
 a system, or matrix model, is solved by the CFT approach, this
 implies that it is solved by the topological recursion on the same
 spectral curve, which is can be more useful for explicit
 computations.

The quasiclassical expansion considered here can be applyed to
systems with   $U(N)$ symmetry, where the
eigenvalues of the matrix variables have fermionic statistics and the
partition function is a $\tau$-function.  In this case the
quasiclassical expansion of the free energy contains only even powers
of $\hbar= 1/N$.  
Our approach
can be carried on also for non-trivial exchange statistics of the
eigenvalues, which is the case for systems with $O(N)$ or $Sp(N)$
symmetry and, more generally, for the $\b$-ensembles with $\b\ne 1$.
In such systems the $1/N$ expansion contains also odd powers in $1/N$.
The first orders were found in \cite{A.Zabrodin:2006aa}, whereas a TR
formalism for hyperelliptical curves was derived in
\cite{Chekhov:2009aa}.  In order to generalize the CFT method to the case
of $\b$-ensembles, it is sufficient to add a term with a background
charge $Q\sim \b -1$ in the definition of the stress-energy tensor
\re{defTz},
\be
T(z) dz^2 =\hf \lim_{z'\to z} \[  d\Phi (z) d\Phi(z') -B(z,z') -  {\b-1\over\sqrt{2\b}}
 d^2\Phi\].
\no
\ee
  This will change the  right state $|\Omega\rangle$ in
the operator representation \re{FockZ}, while the left state $\langle
\Sigma|$ will remain the same.

 Let us make some remarks concerning the application 
 of the  CFT/TR procedure to concrete matrix models.
 The case of the one-matrix model, where the spectral curve is hyperelliptic, 
  is considered in details in \cite{Kostov:2009aa}.
In the case of the $O(n)$ model \cite{Kostov:1988fy}
and the ADE models  \cite{Kostov:1992ie} the spectral curve has a
symmetry relating the branch points on different sheets of the Riemann
surface.   This makes it similar to a hyperelliptic curve.  If such a
symmetry is present, in the definition of the states \re{defsigma}
and \re{defomega}, one should identify all branch points belonging to
the same orbit.

In the case of the two-matrix model with potential $\Tr [ V_1({\bf X}
) + {\bf X} {\bf Y} +V_2({\bf Y})]$, there is a symmetry
$x\leftrightarrow y$ and one can construct a second Fock space
representation of the partition function, based on the conformal
invariance of the spectral plane of the matrix ${\bf Y}$.  The two
representations determine the same partition function.  This
non-trivial fact was proved on the TR side in \cite{Chekhov:2006vd, Eynard:2007ad}.
The first Fock space representation is useful for computing the
correlation functions of the matrix ${\bf X}$, while the second one is
useful for computing the correlation functions of the matrix ${\bf
Y}$.  
In this context a generalised 
 topological recursion procedure was developed  for computing the mixed trace
correlation functions of the matrices ${\bf X}$ and ${\bf Y}$, or equivalently 
free energy in the case when the potential is an arbitrary polynomial
of ${\bf X}$ and ${\bf Y}$
\cite{Eynard:2008af}. The  CFT description of the $1/N$ expansion for
these correlation functions
is still missing.  For that one should construct a
representation of the $W_\infty$ algebra on the complex curve.  This
is an interesting problem which we would like to address in the
future.

  \bigskip \leftline{\bf Acknowledgments}

  \noindent Part of this work has been done during the visit of I.K.
  at the Departamento de Matem\'atica, Instituto Superior T\'ecnico,
  Lisboa.  N.O. thanks B. Eynard for useful discussions and the
  Institut de Physique Th\'eorique, CEA Saclay, for its kind
  hospitality .

\appendix

\section{Solving the Virasoro constraints near a branch point}
\def\gs{ g_{\text{s}}} \la{appendixA}

Here we obtain the solution to the Virasoro constraints (we have put
$\hbar=1$)
\be\la{lmOm1} L_{2n} &=& {1\over 4} \sum_{k+q=2n} :\left(J_k +
\mu_{-k} \right) \left( J_{q}+ \mu _{-q} \right): + {1\over 16
}\delta_{n,0} , \ee
 which determine the coefficients $w^{(g)} _{p_1,\dots , p_n} $.  We
 represent the oscillator amplitudes $J_p$ as
\be\la{ReprJt} \mu_p+ J _{-p} \to t_{p} , \, , \quad J_{p} \to
p{\p\over \p t_{p}} \qquad (p\ge 1).  \ee
Then the Virasoro operators are represented by
the differential operators 
\be\la{Virtn} 4 \hat L_{-2} &= & \sum_{k\ge 1} k t_{k+2} \p_k F +
t_1^2 , \no
\\
4 \hat L_0 &=&2 \sum_{k\ge 1} k t_k \p_k F + {1\over 4}, \no\\
4 \hat L_{p-3} &=& \sum_{p-k-q=3} k\, q \, \p_k\p_q + 2 \sum_{p+k-q=3}
q t_k \p_q \hskip 1cm (p\ge 5, \netwo) \ee
  which act on the partition function $Z[t]\equiv Z[t_1, t_3, \dots]$:
 \be \hat L_{p-3} \ Z [t]= 0, \quad p =1, 3, 5, \dots.  \ee

We are interested in   solution with $t_3\ne 0$.   
 Then the  $n=0$, or $p=3$, constraint  is solved by 
assuming that the partion function is of the form
\be\la{scfrmt}
Z [t_p] =t_3^{-1/24} \ e^{F [ t_p / t_3^{p/3} ]}
. 
\ee
We look for a solution for the free energy $F $, which is
   a formal expansion in $t_1, t_5, t_7,  \dots $
and $1/t_3\sim 1/N$:
  \be\la{expOmo} F = \sum_ {g\ge 0} \ \sum_{n\ge0} {(-1)^n\over n!}
  \sum_{k_1,\dots, k_n\ge 1}{ t _3 } ^{2-2g-n} \ w^{(g)}_{k_1, \dots,
  k_n}\ {t_{k_1}\dots t_{k_n}\over k_1\dots k_n}.  \ee
 The compatibility of this expansion  with   \re{scfrmt}   implies that 
 the coefficients $ w^{(g)}_{k_1, \dots, k_n}$
are nonzero only if 
\be\la{sccb}
\sum _{i=1}^n k_i = 3 (2g-2+n).
\ee

Written for the free energy $F=\ln Z$, , the Virasoro constraints
become a set of quadratic constraints for the derivatives
\be w_p \defeq - p \p_p F, \ \ w_{p, q} \defeq pq \p_p\p_q F. \ee
 We have to solve for $p=1,3,\dots$
\be\la{viras} \sum_{p-k-q=3} \( w_k\, w_q + w_{k,q} \) - { 2 }
\sum_{p+k-q=3} t_k w_q + t_1^2\, \d_{p, 1} =0 .  \ee
 In particular, the equation for $p=3$ means that the generating
 function
\be \phi(z) =- \sum_{p\ge 1} {w_p\over p z^p} + \sum _{p\ge 1} t_p
{z^p\over p} \ee
is scale invariant with respect to 
$t_p\to \rho^{-p}t_p, \ z\to \rho z$.

    One can solve \re{viras} recursively in the genus $g$
    by expanding
 \be
 w_p[t_k] = \sum_{g\ge 1} {t_3}^{ 2-2g} w_p^{(g)}[t_k/t_3].
 \ee
 The solution of \re{viras} can be obtained as a sum of graphically
 represented by introducing a trivalent vertex which gives the
 conservation law of the three incoming momenta $p, q, k$:
 \be
 v_{p, q, k } ={1\over t_3}   \d_{p+q+k-3, 0}.
 \ee
This vertex satisfies the selection rule \re{sccb} with $g=0$ and
$n=3$.  Then \re{viras} takes the form
\be
  w_p^{(g)} = v_{p, -k,-q}  \(\hf  \sum_{h=0} ^g 
  w_k^{(h)} w_q^{(g-h)} + w_{k,q} ^{(g-1)} \) 
+ 2    v_{p, k,-q} t_k w_q^{(g)}  +\hf   v_{p, 1,1}  t_1^2  .
\ee
This is the generating function for a set of recurrence equations for
the coefficients $w^{(g)} _{p_1,\dots , p_n} $.  The equation for the
correlator with $n+1$ legs $p, p_1, \dots , p_n$ is obtained by
differentiating w.r.t. $t_{p_1} , \dots, t_{p_n}$ and then taking $t_q
= \d_{q, 3}$.  The equation has the following graphical
representation:
\bigskip \be\la{recureqb} \begin{array}{c}{
\epsfxsize=10truecm\epsfbox{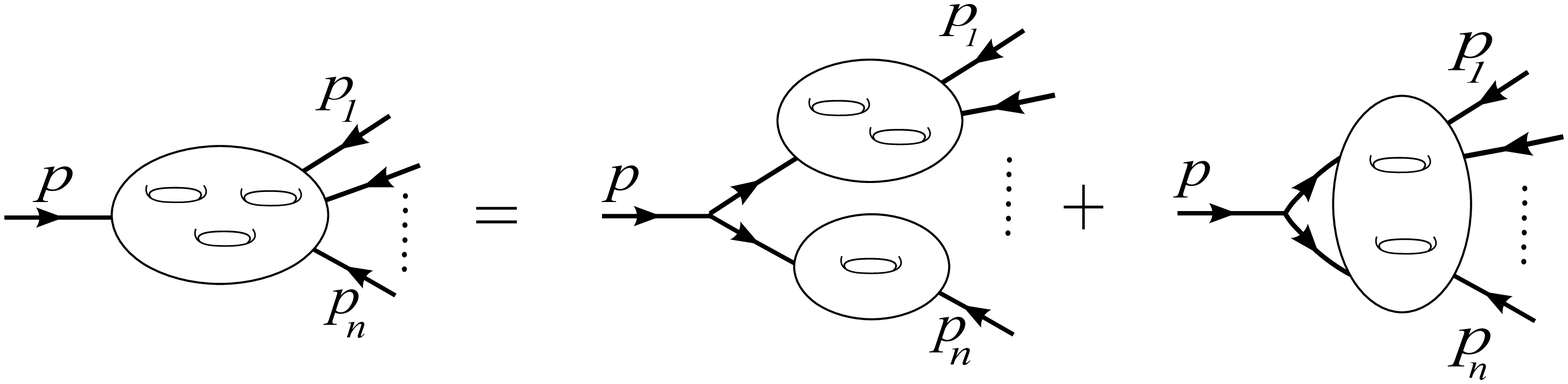} }\end{array}\, \ .  \ee
\bigskip 
 This graphical representation is related to that for the topological
 recursion \re{recureqb} by replacing
\be\la{vertexA}
\begin{array}{c}{

\epsfxsize=5truecm\epsfbox{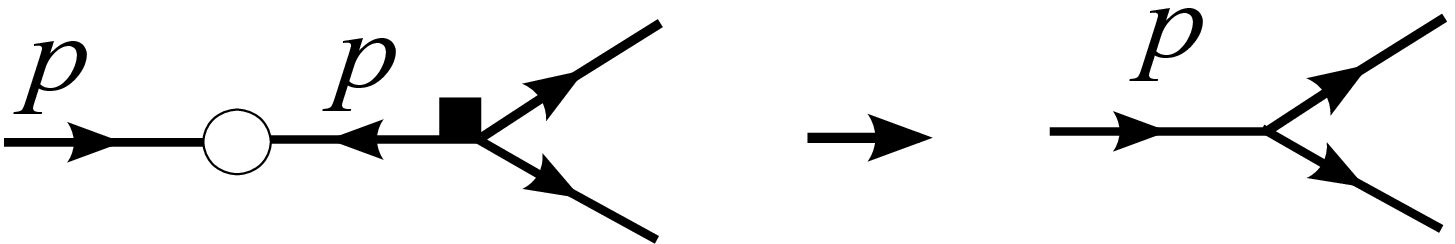} }\end{array}\, \ .  \ee
\bigskip 
%

  \small 
 
 \providecommand{\href}[2]{#2}\begingroup\raggedright\endgroup

    \end{document}